\documentclass[journal,10pt,twocolumn]{IEEEtran}

\usepackage{amsmath,amssymb,amsfonts}
\usepackage{bm}
\usepackage{multirow}
\usepackage{tabularx}
\usepackage{color}
\usepackage{cite}
\usepackage{algorithm,algorithmic}
\usepackage{caption}
\usepackage{subcaption}
\usepackage{blindtext}
\usepackage{lettrine}
\usepackage{float}
\usepackage{makecell}
\usepackage[export]{adjustbox}
\usepackage{graphicx}
\usepackage{booktabs}
\usepackage[dvips]{epsfig}
\usepackage{boxedminipage}
\usepackage{stfloats}
\usepackage{lipsum}

\usepackage{scalerel}
\usepackage{tikz}
\usetikzlibrary{svg.path}

\definecolor{orcidlogocol}{HTML}{A6CE39}
\tikzset{
    orcidlogo/.pic={
        \fill[orcidlogocol] svg{M256,128c0,70.7-57.3,128-128,128C57.3,256,0,198.7,0,128C0,57.3,57.3,0,128,0C198.7,0,256,57.3,256,128z};
        \fill[white] svg{M86.3,186.2H70.9V79.1h15.4v48.4V186.2z}
        svg{M108.9,79.1h41.6c39.6,0,57,28.3,57,53.6c0,27.5-21.5,53.6-56.8,53.6h-41.8V79.1z M124.3,172.4h24.5c34.9,0,42.9-26.5,42.9-39.7c0-21.5-13.7-39.7-43.7-39.7h-23.7V172.4z}
        svg{M88.7,56.8c0,5.5-4.5,10.1-10.1,10.1c-5.6,0-10.1-4.6-10.1-10.1c0-5.6,4.5-10.1,10.1-10.1C84.2,46.7,88.7,51.3,88.7,56.8z};
    }
}

\newcommand\orcidicon[1]{\href{https://orcid.org/#1}{\mbox{\scalerel*{
                \begin{tikzpicture}[yscale=-1,transform shape]
                \pic{orcidlogo};
                \end{tikzpicture}
            }{|}}}}

\usepackage[bookmarks=false]{hyperref}

\DeclareCaptionLabelSeparator{periodspace}{.\quad}
\DeclareMathOperator*{\argmin}{argmin}
\IEEEoverridecommandlockouts

\begin{document}
%
\title{Location-aware Channel Estimation for RIS-aided mmWave MIMO Systems via Atomic Norm Minimization 
\thanks{This research was supported by the MSIP (Ministry of Science, ICT and Future Planning), Korea, under the ITRC (Information Technology Research Center) support program (IITP-2021-2017-0-01637) supervised by the IITP (Institute for Information \& communications Technology Promotion).
\newline \indent Hyeonjin Chung and Sunwoo Kim are with the Department of Electronics and Computer Engineering, Hanyang University, Seoul, 04763, South Korea (email: hyeonjingo@hanyang.ac.kr; remero@hanyang.ac.kr).
}}

\author{
\IEEEauthorblockN{Hyeonjin Chung~\IEEEmembership{Student Member,~IEEE,} and Sunwoo Kim~\IEEEmembership{Senior Member,~IEEE}}}
\maketitle

\begin{abstract}
In this paper, we propose a location-aware channel estimation based on the atomic norm minimization (ANM) for the reconfigurable intelligent surface (RIS)-aided millimeter-wave multiple-input-multiple-output (MIMO) systems.
The beam training overhead at the base station (BS) is reduced by the direct beam steering towards the RIS with the location of the BS and the RIS.
The RIS beamwidth adaptation is proposed to reduce the beam training overhead at the RIS, and also it enables accurate channel estimation by ensuring the user equipment receives all the multipath components from the RIS.
After the beam training, the cascaded effective channel of the RIS-aided MIMO systems is estimated by ANM.
Depending on whether the beam training overhead at the BS or at the RIS is reduced or not, the channel is represented as a linear combination of either 1D atoms, 2D atoms, or 3D atoms, and the ANM is applied to estimate the channel. Simulation results show that the proposed location-aware channel estimation via 2D ANM and 3D ANM achieves superior estimation accuracy to benchmarks. 
\end{abstract}

\begin{IEEEkeywords}
Reconfigurable intelligent surface, atomic norm minimization, multi-dimensional atomic norm, beamwidth adaptation, channel estimation, low-overhead, MIMO
\end{IEEEkeywords}

\IEEEpeerreviewmaketitle

\section{Introduction}
A millimeter-wave (mmWave) has been considered as a solution to accommodate increasing data traffic and frequency spectrum shortage~\cite{6515173}.
Over few years, 5G NR and IEEE 802.11 have worked on a standard for mmWave communications, and the standardization is still in progress up to date~\cite{5Gsurvey,IEEE802}.   
Moreover, a communication on terahertz (THz) frequency band has been pointed out as one of the candidate technologies for 6G, where 6G is expected to support heavy applications such as virtual reality, augmented reality, and holograms~\cite{Samsung}.
However, mmWave and THz signals suffer from high path loss and are susceptible to blockages due to the weak diffraction. To address the high path loss, a massive antenna array is deployed to form a highly directional beam, which can extend the coverage~\cite{6736761}.
On the other hand, the weak diffraction is still a difficult issue to solve since the strength of the mmWave and the THz signal significantly drops if the line-of-sight (LoS) path is blocked.

Recently, a reconfigurable intelligent surface (RIS) has been studied to address the weak diffraction~\cite{9140329}.
Normally, the RIS is made up of either reflectarray or metasurface, and each passive element or unit cell can programmably change the propagation characteristic of the incident wave, such as phase, reflection angle, and refraction angle~\cite{9086766}.
In the seminal works on RIS-aided mmWave and THz communications, the RIS is located in the place where a base station (BS) is visible and focuses the incident wave towards user equipment (UE) via reflect beamforming~\cite{9103231,9314027,8959174}. By this way, the signal can bypass the blockage between the BS and the UE so that the communication becomes more reliable.     
A concept of RIS-aided mmWave or THz communications has generated various new research area, including channel estimation for RIS-aided systems~\cite{9103231}, reflect beamforming design~\cite{9314027}, RIS-aided UAV communications~\cite{8959174}, aerial RIS~\cite{9356531}, and RIS-aided localization~\cite{9215972}.

The channel for RIS-aided systems can be represented as a cascade of two channels: BS-to-RIS channel and RIS-to-UE channel.
Generally, estimating BS-to-RIS channel and RIS-to-UE channel separately is difficult since the RIS only consists of passive elements~\cite{9328501}. Thus, the most of works estimate the cascaded effective channel instead, where the cascaded effective channel denotes a combination between the BS-to-RIS channel and the RIS-to-UE channel.  
To estimate the cascaded effective channel for RIS-aided systems, the BS, the RIS, and the UE have to perform beam search over the entire angular domain, and this results in excessive beam training overhead. 
For this reason, the large beam training overhead has been considered as one of major issues of the channel estimation for RIS-aided systems. 

The algorithm in~\cite{9400843} utilizes the property of the BS-to-RIS channel and the RIS-to-UE channel to reduce the beam training overhead, where the BS-to-RIS channel is static compared to the RIS-to-UE channel. With the BS that supports the full-duplex operation, the static BS-to-RIS channel is estimated less frequently than the RIS-to-UE channel. 
In~\cite{9103231,9354904}, the channel estimation algorithms based on the compressive sensing are proposed. Without excessive beam training overhead, these algorithms can estimate channel parameters such as angle-of-departures (AoDs), angle-of-arrivals (AoAs), and channel gains.
However, the channel estimation accuracy of \cite{9103231,9354904} is limited by a grid-mismatch problem~\cite{5710590}.
To enhance the channel estimation accuracy, the channel estimation algorithm based on the 1D atomic norm minimization (ANM) is proposed in~\cite{he2021channel}, where the 1D ANM is free from the grid-mismatch problem~\cite{6576276}. Here, BS-to-RIS AoDs and RIS-to-UE AoAs are estimated by 1D ANM. Then, the remaining channel parameters are estimated.
In~\cite{9382000}, the channel estimation based on the location information and the 1D ANM has been proposed. Assuming the LoS path always exists, and the location of the BS, the RIS, the UE, and the objects surrounding UE are given, the algorithm in~\cite{9382000} reduces the beam training overhead by restricting the angular domain to search. The AoDs/AoAs are estimated via 1D ANM after the beam training.

Still, there are some issues that have not been considered and addressed in the related works. 
Unlike~\cite{9382000}, the LoS path between the RIS and the UE may not exist when the UE hides behind the structure, and obtaining the location of objects surrounding UE may not be feasible in practice. 
Also, if the beam training overhead at the RIS is reduced, the UE may not receive some multipath components from the RIS. Although this issue makes the channel estimation inaccurate, it is not properly discussed in~\cite{9103231,9354904,he2021channel}, and its solution has not been proposed.
In~\cite{he2021channel,9382000}, the 1D ANM is employed to estimate the AoDs/AoAs, however, the cascaded effective channel can be directly estimated if the atoms and the atomic set are modeled with proper dimension.
In this paper, we propose a location-aware channel estimation based on the ANM, which addresses aforementioned issues. Main distinctions and contributions of this paper are summarized as follows: 
\begin{itemize}
  \item   
  The proposed algorithm exploits the location of the BS and the RIS to reduce the beam training overhead. During the beam training, the BS directly steers the beam towards the RIS.
  \item 
  When the beam training overhead at the RIS is reduced, a proposed RIS beamwidth adaptation widens a reflect beam created by RIS to ensure the UE receives all the multipath components from the RIS.
  \item
  The proposed algorithm estimates the cascaded effective channel via multi-dimensional ANM~\cite{7451201}. When the BS steers the beam towards the RIS using location information, we reveal that the channel is represented as a linear combination of 2D atoms or 3D atoms. 
\end{itemize}

The rest of the paper is organized as follows. Section~\ref{system model} defines the channel model for the downlink RIS-aided systems. Section~\ref{BTSec} presents the beam training procedures for both non location-aware and location-aware scenarios. Section~\ref{RISBWA} introduces the RIS beamwidth adaptation which reduces beam training overhead at the RIS. Section~\ref{alg} introduces the atomic norm and proposes non location-aware channel estimation and location-aware channel estimation which are based on ANM.
Section~\ref{simulation} provides the simulation results and the analysis, and Section~\ref{conclusion} concludes this paper.

$\textit{Notations:}$ We use lower-case and upper-case bold characters to respectively represent vectors and matrices throughout this paper. $(\cdot)^{T}$, $(\cdot)^{H}$, and $(\cdot)^{*}$ respectively denote transpose, conjugate transpose, and complex conjugation.
$(\cdot)^{-1}$ denotes the inverse of a matrix. $\textrm{Tr}(\cdot)$ denotes the trace of a matrix, and $\textrm{diag}(\cdot)$ denotes the diagonal matrix whose diagonal entries equal to entries of given vector. $\textrm{vec}(\cdot)$ denotes vectorization of given matrix.
$\lVert \cdot \rVert_{2}$ and $\lVert \cdot \rVert_{\textrm{F}}$ respectively denote L2 norm and Frobenius norm. 
The curled inequality symbol $\succeq$ denotes matrix inequality. If $\mathbf{A} \succeq \mathbf{B}$, a matrix $\mathbf{A}-\mathbf{B}$ is positive semidefinite.
$\otimes$ and $\diamond$ respectively denote Kronecker product and Khatri-Rao product.
$\mathbf{0}_{N}$ and $\mathbf{I}_{N}$ respectively denote a $N \times 1$ zero vector and a $N \times N$ identity matrix. $\mathcal{CN}(\bm{\mu},\bm{\Sigma})$ denotes a circularly-symmetric complex Gaussian distribution whose mean is $\bm{\mu}$ and covariance is $\bm{\Sigma}$. 

\section{Signal Model}\label{system model}
We consider a downlink RIS-aided mmWave multiple-input-multiple-output (MIMO) system, which means that the BS transmits the signal to the RIS and the RIS bounces back the signal to the UE.
The BS, the RIS, and the UE are equipped with $M_{\textrm{B}}$, $M_{\textrm{R}}$, and $M_{\textrm{U}}$ antennas respectively. Here, antenna arrays that BS, RIS, and UE use are uniform linear arrays (ULAs) with half-wavelength spacing. 
In this paper, the BS and the UE employ full-complexity hybrid beamforming structure~\cite{hybrid}, where the BS and the UE are respectively equipped with $N_{\textrm{B}}$ and $N_{\textrm{U}}$ RF chains. 
Owing to the property of the full-complexity hybrid beamforming, the BS and the UE can simultaneously form up to $N_{\textrm{B}}$ beams and $N_{\textrm{U}}$ beams.   

\begin{figure}[!t]
    \begin{center}
    \includegraphics[width=0.95\columnwidth]{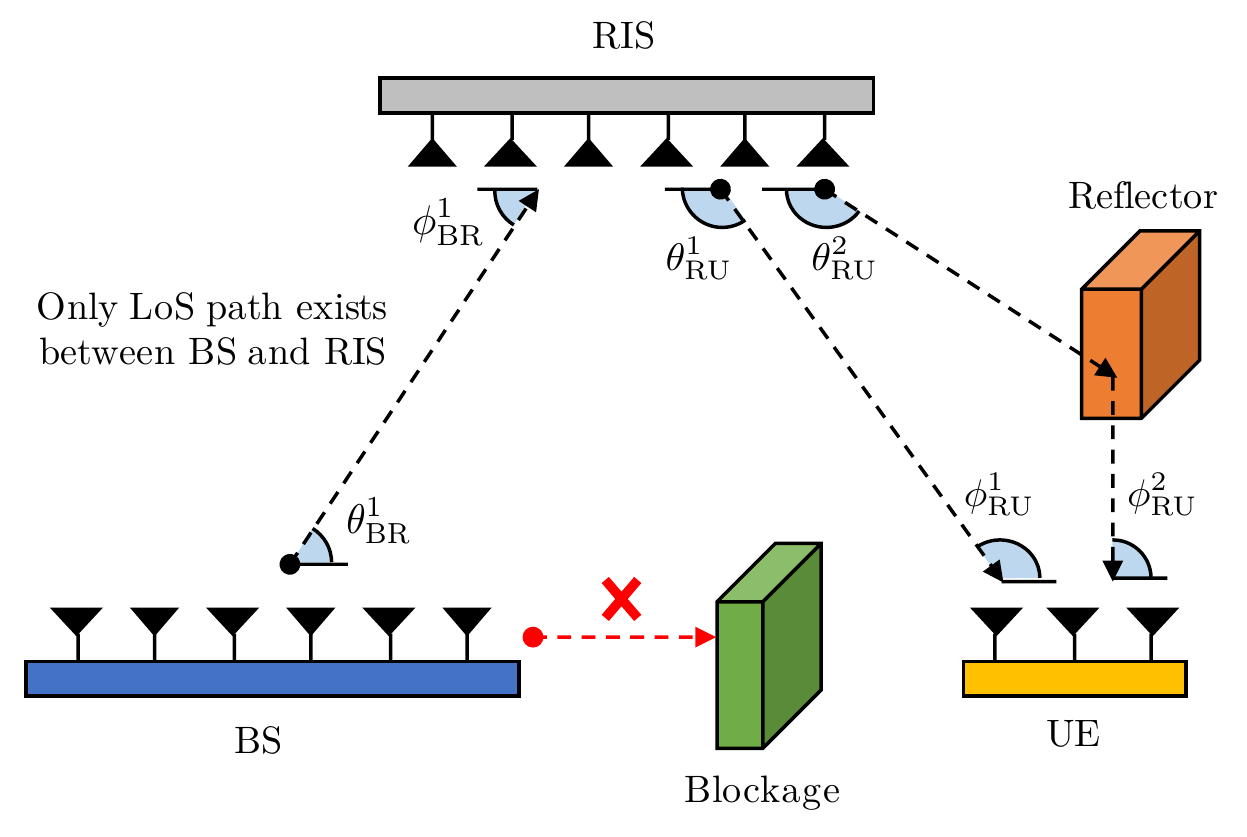}
    \caption{A scheme of RIS-aided mmWave MIMO system and propagation paths. The signal path between BS and UE is blocked.}
    \label{RIS}
    \end{center}
\end{figure}

The steering vector of the ULA with half-wavelength spacing, $\mathbf{a}(\theta)$ is
\begin{equation}\label{steer_vec}
    \mathbf{a}(\theta) = [1,e^{j\pi\cos\theta},\ldots,e^{j\pi(M-1)\cos\theta}]^T \in \mathbb{C}^{M\times 1},
\end{equation}
where $\theta$ denotes the steering direction, and $M$ denotes the number of antennas.
A scheme of RIS-aided mmWave MIMO system and propagation paths is given in Fig.~\ref{RIS}.
Assuming all signal paths between BS and UE are blocked, a channel for RIS-aided mmWave MIMO system can be represented as a cascade of two separate channels: BS-to-RIS channel and RIS-to-UE channel.
The BS-to-RIS channel $\mathbf{H}_{\textrm{BR}}$ can be given by
\begin{equation}\label{HBR}
    \begin{split}
    \mathbf{H}_{\textrm{BR}} &=\sum_{l=1}^{L_{\textrm{BR}}} \alpha_{\textrm{BR}}^{l} \mathbf{a}(\phi_{\textrm{BR}}^{l}) \mathbf{a}(\theta_{\textrm{BR}}^{l})^{H} \\
    &= \mathbf{A}(\bm{\phi}_{\textrm{BR}}) \textrm{diag}(\bm{\rho}_{\textrm{BR}}) \mathbf{A}(\bm{\theta}_{\textrm{BR}})^{H} \in \mathbb{C}^{M_{\textrm{R}} \times M_{\textrm{B}}},
    \end{split}
\end{equation}
where $L_{\textrm{BR}}$ denotes the number of signal paths between BS and RIS. $\alpha_{\textrm{BR}}^{l}$, $\phi_{\textrm{BR}}^{l}$, and $\theta_{\textrm{BR}}^{l}$ respectively denote the channel gain, the BS-to-RIS AoA, and the BS-to-RIS AoD of the $l$-th signal path. 
$\bm{\phi}_{\textrm{BR}}= \{ {\phi}^{1}_{\textrm{BR}},\ldots,{\phi}^{L_{\textrm{BR}}}_{\textrm{BR}} \}$ and $\bm{\theta}_{\textrm{BR}}= \{ {\theta}^{1}_{\textrm{BR}},\ldots,{\theta}^{L_{\textrm{BR}}}_{\textrm{BR}} \}$.
$\mathbf{A}(\bm{\phi}_{\textrm{BR}})=[\mathbf{a}(\phi_{\textrm{BR}}^{1}),\ldots,\mathbf{a}(\phi_{\textrm{BR}}^{L_{\textrm{BR}}}) ] \in \mathbb{C}^{M_{\textrm{R}} \times L_{\textrm{BR}}}$, $\mathbf{A}(\bm{\theta}_{\textrm{BR}})=[\mathbf{a}(\theta_{\textrm{BR}}^{1}),\ldots,\mathbf{a}(\theta_{\textrm{BR}}^{L_{\textrm{BR}}}) ] \in \mathbb{C}^{M_{\textrm{B}} \times L_{\textrm{BR}}}$, and $\bm{\rho}_{\textrm{BR}}=[\alpha_{\textrm{BR}}^{1},\ldots,\alpha_{\textrm{BR}}^{L_{\textrm{BR}}}]^{T}$. 
To bypass the blockage between BS and UE, we assume that the LoS path between BS and RIS always exists. Also, BS and RIS are assumed to be placed where there is no reflector or scatterer nearby.  
Thus, considering the surrounding environment of BS and RIS and sparse propagation characteristic of mmWave~\cite{7501500}, there is only LoS path between BS and RIS so that $L_{\textrm{BR}}=1$. 

The RIS-to-UE channel $\mathbf{H}_{\textrm{RU}}$ can be given by
\begin{equation}\label{HRU}
    \begin{split}
    \mathbf{H}_{\textrm{RU}} &=\sum_{l=1}^{L_{\textrm{RU}}} \alpha_{\textrm{RU}}^{l} \mathbf{a}(\phi_{\textrm{RU}}^{l}) \mathbf{a}(\theta_{\textrm{RU}}^{l})^{H} \\
    &= \mathbf{A}(\bm{\phi}_{\textrm{RU}}) \textrm{diag}(\bm{\rho}_{\textrm{RU}}) \mathbf{A}(\bm{\theta}_{\textrm{RU}})^{H} \in \mathbb{C}^{M_{\textrm{U}} \times M_{\textrm{R}}},
    \end{split}
\end{equation}
where $L_{\textrm{RU}}$ denotes the number of signal paths between RIS and UE. $\alpha_{\textrm{RU}}^{l}$, $\phi_{\textrm{RU}}^{l}$, and $\theta_{\textrm{RU}}^{l}$ respectively denote the channel gain, the RIS-to-UE AoA, and the RIS-to-UE AoD of the $l$-th signal path.
$\bm{\phi}_{\textrm{RU}}= \{ {\phi}^{1}_{\textrm{RU}},\ldots,{\phi}^{L_{\textrm{RU}}}_{\textrm{RU}} \}$ and $\bm{\theta}_{\textrm{RU}}= \{ {\theta}^{1}_{\textrm{RU}},\ldots,{\theta}^{L_{\textrm{RU}}}_{\textrm{RU}} \}$.
$\mathbf{A}(\bm{\phi}_{\textrm{RU}})=[\mathbf{a}(\phi_{\textrm{RU}}^{1}),\ldots,\mathbf{a}(\phi_{\textrm{RU}}^{L_{\textrm{RU}}}) ] \in \mathbb{C}^{M_{\textrm{U}} \times L_{\textrm{RU}}}$, $\mathbf{A}(\bm{\theta}_{\textrm{RU}})=[\mathbf{a}(\theta_{\textrm{RU}}^{1}),\ldots,\mathbf{a}(\theta_{\textrm{RU}}^{L_{\textrm{RU}}}) ] \in \mathbb{C}^{M_{\textrm{R}} \times L_{\textrm{RU}}}$, and $\bm{\rho}_{\textrm{RU}}=[\alpha_{\textrm{RU}}^{1},\ldots,\alpha_{\textrm{RU}}^{L_{\textrm{RU}}}]^{T}$.
Unlike the assumptions related to the BS-to-RIS channel, we assume that there can be multiple signal paths between RIS and UE since the UE is mobile and can be located nearby reflectors or scatterers.
Also, the LoS path between RIS and UE may not exist when the UE is located behind the blockage.
Due to these reasons, $L_\textrm{RU}$ can be larger than $1$, where $L_\textrm{RU}$ depends on the surrounding environment.

A RIS control matrix $\bm{\Omega}$ can be given by
\begin{equation}\label{Omega}
    \bm{\Omega}=
    \begin{bmatrix}
    \beta_{1} e^{j \vartheta_{1}} & 0 & \ldots & 0 \\
    0 & \beta_{2} e^{j \vartheta_{2}} & \ldots & 0 \\
    \vdots & \vdots & \ddots & \vdots \\
    0 & 0 & \ldots & \beta_{M_{\textrm{R}}} e^{j \vartheta_{M_{\textrm{R}}}}
    \end{bmatrix} \in \mathbb{C}^{M_{\textrm{R}} \times M_{\textrm{R}}},
\end{equation}
where $\beta_{m}$ and $\vartheta_{m}$ respectively denote a reflection coefficient and a phase shift of the $m$-th antenna in RIS.
$\vartheta_{m} \in [0,2\pi)$ and $\beta_{m}$ can be either $0$ or $1$, where $0$ and $1$ respectively denotes the deactivation and the activation of the $m$-th antenna in the RIS.  
For a convenient representation of $\bm{\Omega}$, a RIS control vector $\bm{\omega}$ is defined as
\begin{equation}\label{vecOmega}
    \bm{\omega}=\left[\beta_{1} e^{j \vartheta_{1}},\beta_{2} e^{j \vartheta_{2}},\ldots,\beta_{M_{\textrm{R}}} e^{j \vartheta_{M_{\textrm{R}}}} \right]^{T} \in \mathbb{C}^{M_{\textrm{R}} \times 1}.
\end{equation}
Note that $\bm{\Omega}=\textrm{diag}(\bm{\omega})$. The cascaded channel for RIS-aided MIMO system, $\mathbf{H}$ can be given by
\begin{equation}\label{Cascade}
    \mathbf{H}=\mathbf{H}_{\textrm{RU}} \bm{\Omega} \mathbf{H}_{\textrm{BR}} \in \mathbb{C}^{M_{\textrm{U}} \times M_{\textrm{B}}}.
\end{equation}

\section{Beam Training Procedure for RIS-aided MIMO Systems}\label{BTSec}
\begin{figure*}[!t]
    \begin{center}
    \captionsetup[subfigure]{justification=centering}
    \begin{subfigure}[t]{0.82\columnwidth}
        \includegraphics[width=\columnwidth,center]{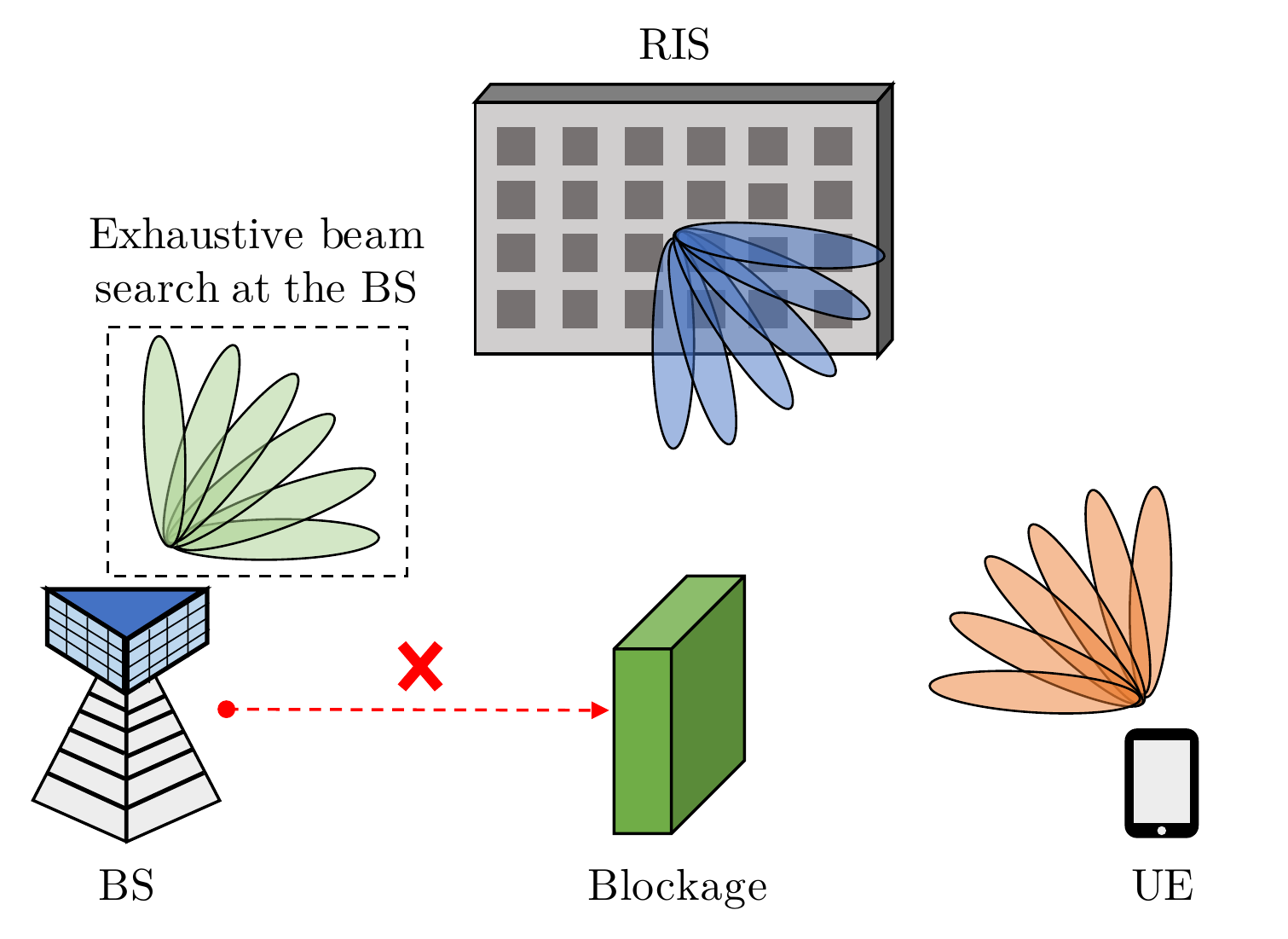}
        \caption{Non location-aware beam training}\label{nla}
    \end{subfigure}
    \begin{subfigure}[t]{0.82\columnwidth}
        \includegraphics[width=\columnwidth,center]{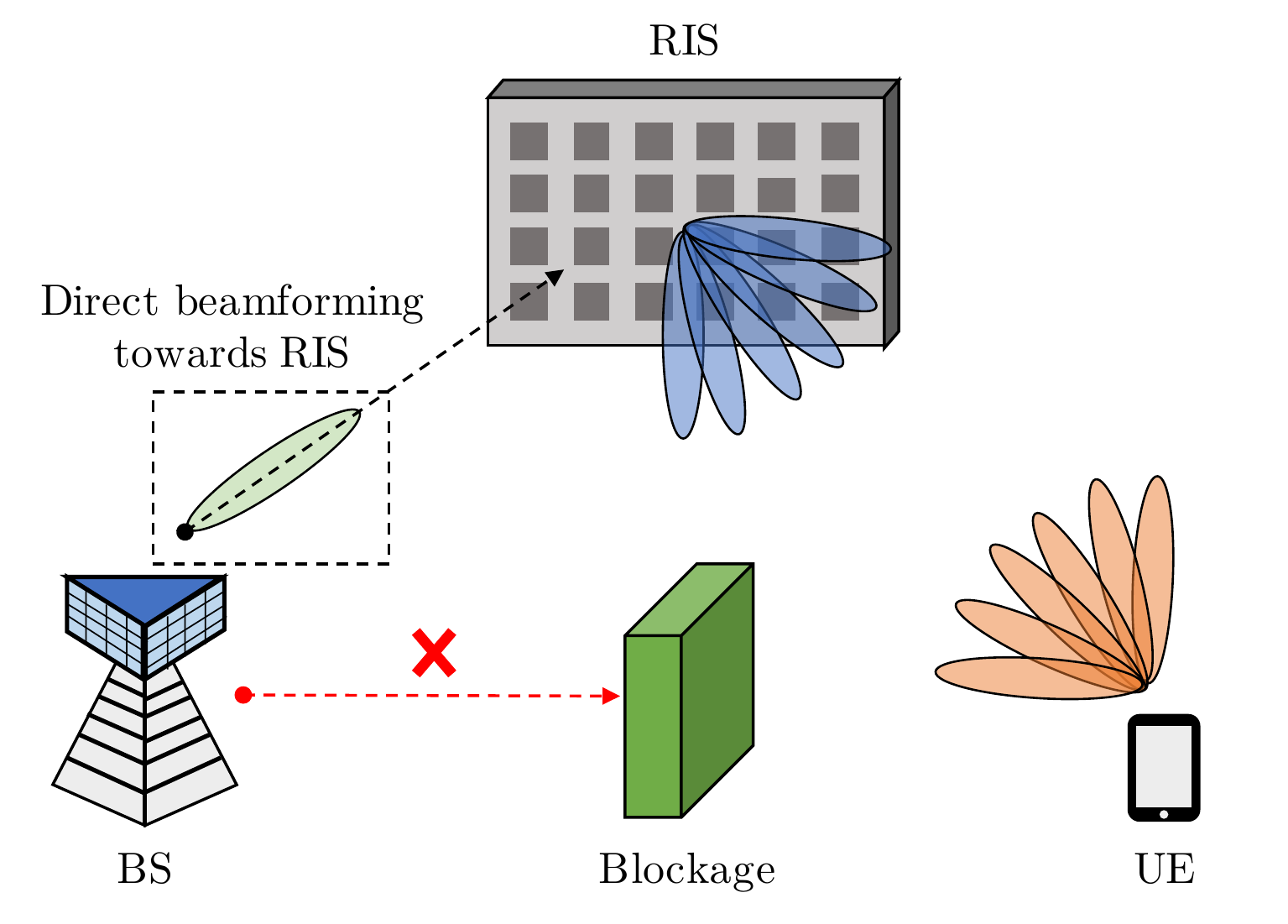}
        \caption{Location-aware beam training}\label{la}
    \end{subfigure}
    \caption{Schemes of non location-aware beam training and location-aware beam training.}
    \label{locationBT}
    \end{center}
\end{figure*}

In this section, we introduce the location-aware beam training whose overhead is reduced by the location of the BS and the RIS. 
Schemes of non location-aware beam training and location-aware beam training are compared in Fig.~\ref{locationBT}. 
Fig.~\ref{nla} shows the procedure of the non location-aware beam training. Since the BS does not know the BS-to-RIS AoD, the BS must perform exhaustive beam search over entire angular domain.  
If the location information is given, the BS-to-RIS AoD can be calculated so that the BS can form a beam towards RIS as in Fig.~\ref{la}.

After the beam training, the received pilot signals collected during the beam training are organized for the channel estimation. Key properties for the organization are given as follows. 
\begin{itemize}
	\item Property 1: $\textrm{vec}\left(\mathbf{A \textrm{diag}(\mathbf{b}) C} \right)=(\mathbf{C}^{T} \diamond \mathbf{A})\mathbf{b}.$
	\item Property 2: $(\mathbf{AB} \diamond \mathbf{CD})=(\mathbf{A} \otimes \mathbf{C})(\mathbf{B} \diamond \mathbf{D}).$
	\item Property 3: $\textrm{vec}\left(\mathbf{A B C} \right)=(\mathbf{C}^{T} \otimes \mathbf{A}) \textrm{vec}(\mathbf{B}).$
\end{itemize}
Definitions and properties of Kronecker product and Khatri-Rao product are well-explained in~\cite{KR}.


\subsection{Non Location-aware Beam Training}\label{NLBT}
\begin{figure}[!t]
    \captionsetup[subfigure]{justification=centering}
    \begin{subfigure}[t]{1\columnwidth}
    \includegraphics[width=1\columnwidth]{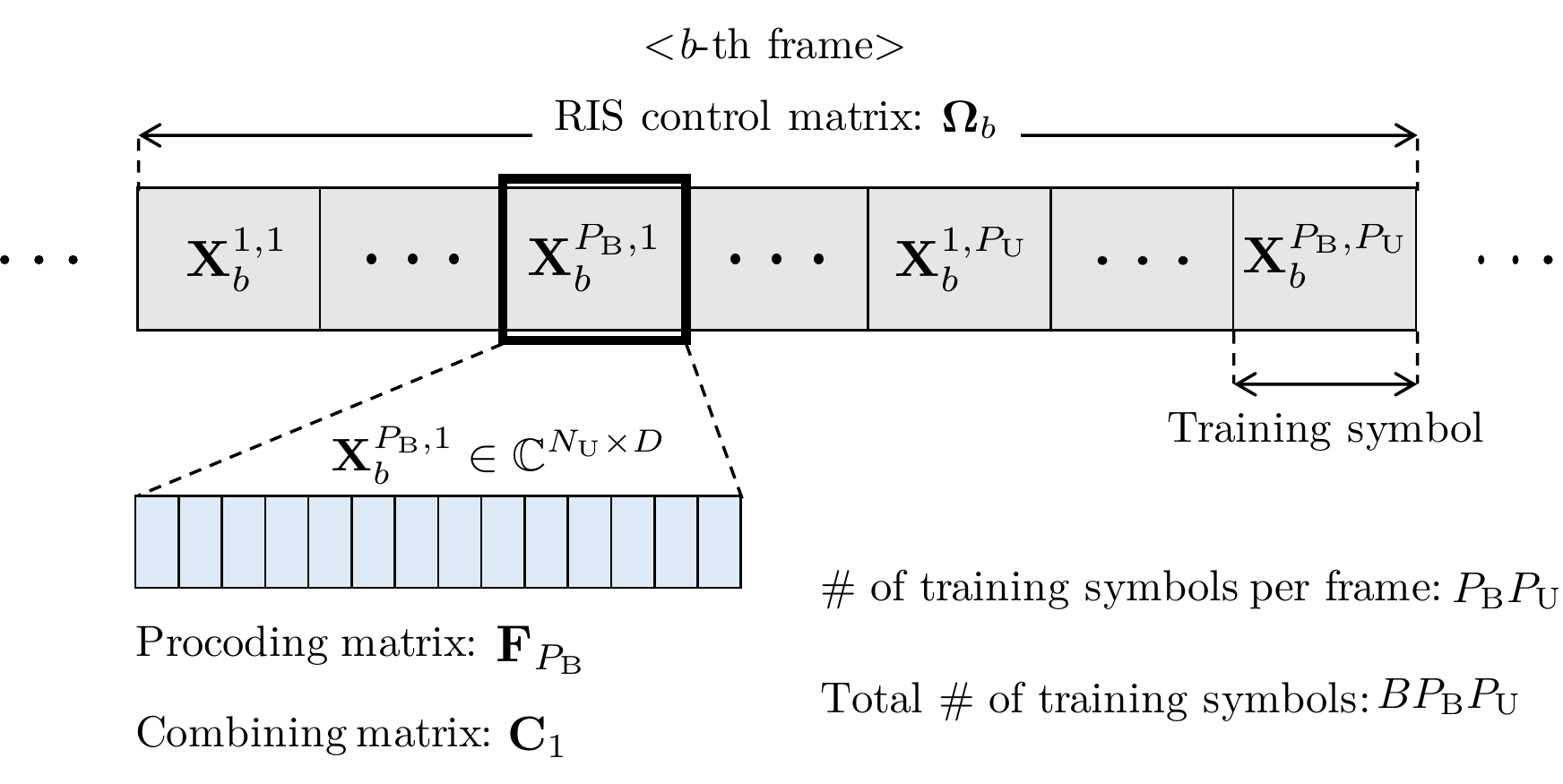}
    \caption{Non location-aware beam training}\label{Fig1}
    \end{subfigure}
    \begin{subfigure}[t]{1\columnwidth}
    \includegraphics[width=1\columnwidth]{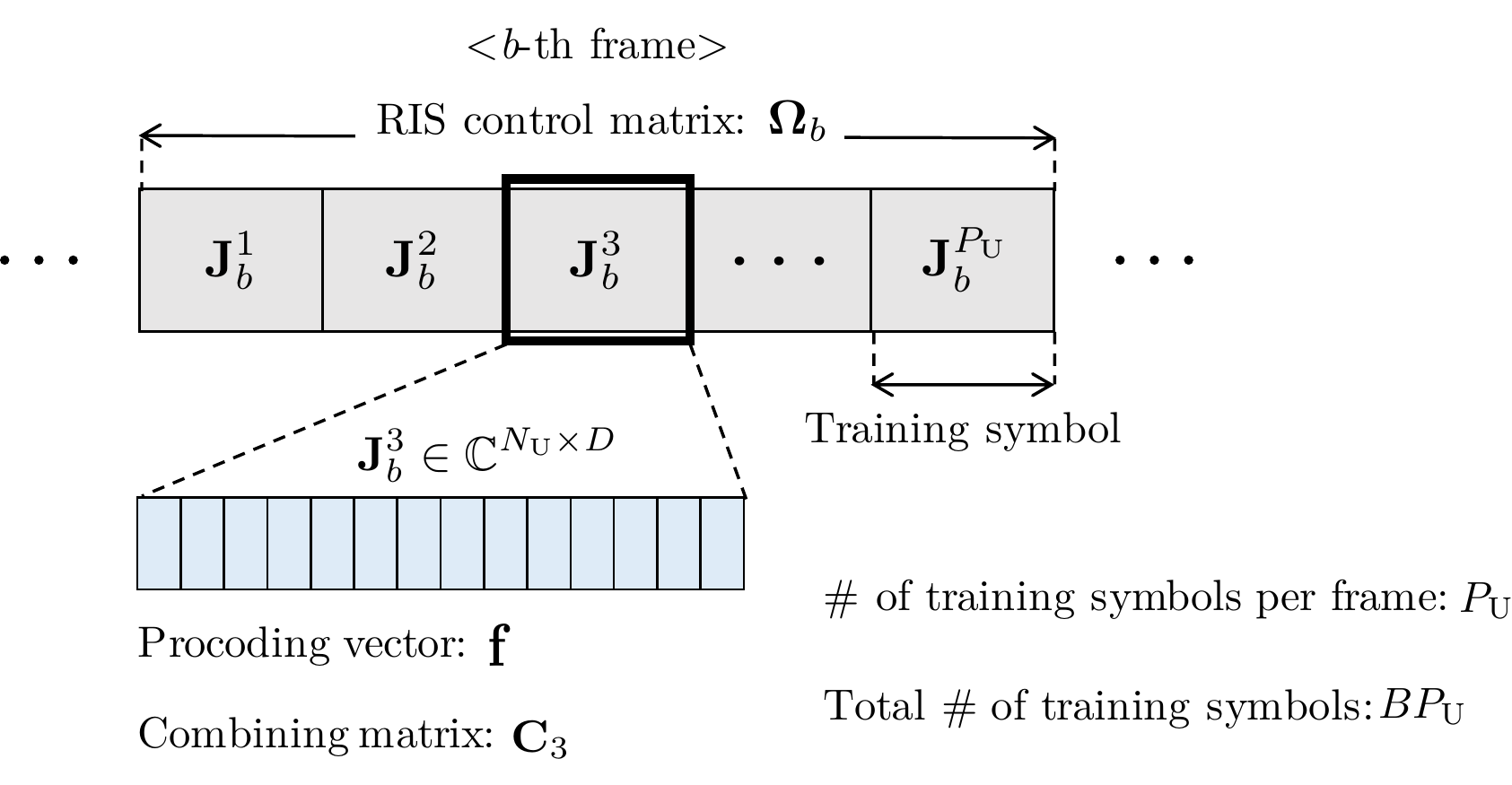}
    \caption{Location-aware beam training}\label{Fig1_v2}
    \end{subfigure}
    \caption{A frame structure for non location-aware beam training and location-aware beam training in RIS-aided MIMO communication.}
\end{figure}

A frame structure for non location-aware beam training in RIS-aided MIMO systems is depicted in Fig.~\ref{Fig1}.
To simplify notations, $M_{\textrm{B}}/N_{\textrm{B}}$ and $M_{\textrm{U}}/N_{\textrm{U}}$ are respectively defined as $P_{\textrm{B}}$ and $P_{\textrm{U}}$.
Each frame consists of multiple training symbols, and the RIS control matrix changes frame by frame.
There are $P_{\textrm{B}}$ precoding matrices and $P_{\textrm{U}}$ combining matrices, and $\mathbf{F}_{i} \in \mathbb{C}^{M_{\textrm{B}} \times N_{\textrm{B}}}$ and $\mathbf{C}_{j} \in \mathbb{C}^{M_{\textrm{U}} \times N_{\textrm{U}}}$ respectively denote the $i$-th precoding matrix and the $j$-th combining matrix.
The precoding matrix and the combining matrix change by each training symbol, so that one frame consists of $P_{\textrm{B}}P_{\textrm{U}}$ training symbols.
Here, the L2 norm of each precoding vector and combining vector is $1$.
Letting $B$ denotes the number of frames, the total number of training symbols equals to $B P_{\textrm{B}}P_{\textrm{U}}$. 

$\mathbf{X}^{i,j}_{b}$, a received pilot signal at the $b$-th frame which uses the $i$-th precoding matrix and the $j$-th combining matrix, can be given by 
\begin{equation}\label{rxsignal1}
    \mathbf{X}^{i,j}_{b}=\mathbf{C}_{j}^{H} \mathbf{H}_{\textrm{RU}} \bm{\Omega}_{b} \mathbf{H}_{\textrm{BR}} \mathbf{F}_{i} \mathbf{S} + \mathbf{N}^{i,j}_{b} \in \mathbb{C}^{N_{\textrm{U}} \times D},
\end{equation}
where $\bm{\Omega}_{b}$ denotes the RIS control matrix at the $b$-th frame, and $D$ denotes the number of signal samples per one training symbol. $\mathbf{S}=\left[\mathbf{s}_{1},\ldots,\mathbf{s}_{N_{\textrm{B}}} \right]^{T} \in \mathbb{C}^{N_{\textrm{B}} \times D}$, where $\mathbf{s}_{n}$ is the $n$-th pilot signal that satisfies $\mathbf{s}^{H}_{n}\mathbf{s}_{n}/D=P_{\textrm{Tx}}/ N_{\textrm{B}}$.
Here, $P_{\textrm{Tx}}$ denotes the transmitting power of the BS, which is equivalent to $\lVert \mathbf{S} \rVert^{2}_{\textrm{F}}/D$.
The pilot signals are orthogonal to each other so that $\mathbf{S}\mathbf{S}^{H}/D=P_{\textrm{Tx}} \mathbf{I}_{N_{\textrm{B}}}/N_{\textrm{B}}$. 
$\mathbf{N}^{i,j}_{b} \in \mathbb{C}^{N_{\textrm{U}} \times D}$ is a noise matrix whose columns follow $\mathcal{CN}(\mathbf{0}_{N_{\textrm{U}}},\sigma^{2} \mathbf{I}_{N_{\textrm{U}}})$, where $\sigma^{2}$ denotes the power of the noise.

After receiving $\mathbf{X}^{i,j}_{b}$, $\mathbf{S}^{H}$ are multiplied to $\mathbf{X}^{i,j}_{b}$ to filter the noise. 
We define this filtered signal $\mathbf{Y}^{i,j}_{b}$ as   
\begin{equation}\label{rxsignal2}
    \begin{split}
    \mathbf{Y}^{i,j}_{b}&=\frac{\mathbf{X}^{i,j}_{b}\mathbf{S}^{H}}{D}\\
    &=\frac{P_{\textrm{Tx}}}{N_{\textrm{B}}} \mathbf{C}_{j}^{H} \mathbf{H}_{\textrm{RU}} \bm{\Omega}_{b} \mathbf{H}_{\textrm{BR}} \mathbf{F}_{i} + \frac{\mathbf{N}^{i,j}_{b}\mathbf{S}^{H}}{D} \in \mathbb{C}^{N_{\textrm{U}} \times N_{\textrm{B}}}.
    \end{split}
\end{equation}
A total $P_{\textrm{B}}P_{\textrm{U}}$ filtered signals received at the $b$-th frame is organized as
\begin{equation}\label{merged}
    \mathbf{Y}_{b}= 
    \begin{bmatrix}
    \mathbf{Y}^{1,1}_{b} & \mathbf{Y}^{2,1}_{b} & \cdots & \mathbf{Y}^{P_{\textrm{B}},1}_{b} \\
    \mathbf{Y}^{1,2}_{b} & \mathbf{Y}^{2,2}_{b} & \cdots & \mathbf{Y}^{P_{\textrm{B}},2}_{b} \\
    \vdots & \vdots & \cdots & \vdots \\
    \mathbf{Y}^{1,P_{\textrm{U}}}_{b} & \mathbf{Y}^{2,P_{\textrm{U}}}_{b} & \cdots & \mathbf{Y}^{P_{\textrm{B}},P_{\textrm{U}}}_{b}
    \end{bmatrix}
    \in \mathbb{C}^{M_{\textrm{U}} \times M_{\textrm{B}}},
\end{equation}
where $\mathbf{Y}_{b}$ is a compilation of all filtered signals at the $b$-th frame. $\mathbf{Y}_{b}$ can also be represented as
\begin{equation}\label{merged2}
    \mathbf{Y}_{b}=\frac{P_{\textrm{Tx}}}{N_{\textrm{B}}} \mathbf{C}^{H} \mathbf{H}_{\textrm{RU}} \bm{\Omega}_{b} \mathbf{H}_{\textrm{BR}} \mathbf{F} + \mathbf{V}_{b},
\end{equation}
where $\mathbf{F} \in \mathbb{C}^{M_{\textrm{B}} \times M_{\textrm{B}}}$ and $\mathbf{C} \in \mathbb{C}^{M_{\textrm{U}} \times M_{\textrm{U}}}$ respectively denote a full-rank precoding matrix and a full-rank combining matrix. 
$\mathbf{V}_{b} \in \mathbb{C}^{M_{\textrm{U}} \times M_{\textrm{B}}}$ is a matrix that represents the remaining noise, and each column of $\mathbf{V}_{b}$ follows $\mathcal{CN}\left( \mathbf{0}_{M_{\textrm{U}}},(\sigma^{2}P_{\textrm{Tx}}/DN_{\textrm{B}})\mathbf{I}_{M_{\textrm{U}}} \right)$.  

For the channel estimation, $\mathbf{Y}_{b}$ for $b=1,\ldots,B$ should be organized to exploit the pilot signals from all directions. 
Letting a lengthy column vector $\mathbf{y}_{b}$ equals $\textrm{vec}(\mathbf{Y}_{b})$, $\mathbf{y}_{b}$ can be represented as follows by using Property 1.
\begin{equation}
    \begin{split}
    \mathbf{y}_{b}&=\textrm{vec}(\mathbf{Y}_{b})\\
    &=\frac{P_{\textrm{Tx}}}{N_{\textrm{B}}} \left(\mathbf{F}^{T} \mathbf{H}^{T}_{\textrm{BR}} \diamond \mathbf{C}^{H} \mathbf{H}_{\textrm{RU}}\right) \bm{\omega}_{b}  + \mathbf{v}_{b} \in \mathbb{C}^{M_{\textrm{B}}M_{\textrm{U}} \times 1},
    \end{split}
\end{equation}
where $\bm{\omega}_{b}$ denotes the RIS control vector at the $b$-th frame, and $\bm{\Omega}_{b}=\textrm{diag}(\bm{\omega}_{b})$. $\mathbf{v}_{b}=\textrm{vec}(\mathbf{V}_{b})$, where $\mathbf{v}_{b}$ follows $\mathcal{CN}\left(\mathbf{0}_{M_{\textrm{B}}M_{\textrm{U}}},(\sigma^{2}P_{\textrm{Tx}}/DN_{\textrm{B}})\mathbf{I}_{M_{\textrm{B}}M_{\textrm{U}}}\right)$.
Then, we form an organized matrix $\bm{\mathcal{Y}}$ which is constructed by stacking $\mathbf{y}_{b}$ for $b=1,\ldots,B$ as follows.
\begin{equation}
    \begin{split}
    \bm{\mathcal{Y}}&=\left[\mathbf{y}_{1},\ldots,\mathbf{y}_{B} \right]\\&=\frac{P_{\textrm{Tx}}}{N_{\textrm{B}}} \left(\mathbf{F}^{T} \mathbf{H}^{T}_{\textrm{BR}} \diamond \mathbf{C}^{H} \mathbf{H}_{\textrm{RU}}\right) \mathbf{W}  + \bm{\mathcal{V}} \in \mathbb{C}^{M_{\textrm{B}}M_{\textrm{U}} \times B},
    \end{split}
\end{equation}
where $\mathbf{W}=[\bm{\omega}_{1},\ldots,\bm{\omega}_{B}] \in \mathbb{C}^{M_{\textrm{R}} \times B}$ and $\bm{\mathcal{V}}=[\mathbf{v}_{1},\ldots,\mathbf{v}_{B}] \in \mathbb{C}^{M_{\textrm{B}}M_{\textrm{U}} \times B}$. 
By using Property 2, $\bm{\mathcal{Y}}$ can be also represented as
\begin{equation}
    \bm{\mathcal{Y}}=\frac{P_{\textrm{Tx}}}{N_{\textrm{B}}} \left(\mathbf{F}^{T} \otimes \mathbf{C}^{H}\right) \left( \mathbf{H}^{T}_{\textrm{BR}} \diamond  \mathbf{H}_{\textrm{RU}} \right) \mathbf{W}  + \bm{\mathcal{V}}.
\end{equation}
Here, $\mathbf{H}^{T}_{\textrm{BR}} \diamond  \mathbf{H}_{\textrm{RU}} \in \mathbb{C}^{M_{\textrm{B}}M_{\textrm{U}} \times M_{\textrm{R}}}$ contains a channel information that is independent of $\mathbf{F}$, $\mathbf{C}$, and $\mathbf{W}$. 
Once $\mathbf{H}^{T}_{\textrm{BR}} \diamond  \mathbf{H}_{\textrm{RU}}$ is successfully estimated, the optimal RIS control matrix that maximizes SNR can be derived by conducting singular value decomposition (SVD) to $\mathbf{H}^{T}_{\textrm{BR}} \diamond  \mathbf{H}_{\textrm{RU}}$~\cite{he2021channel}. 
Throughout this paper, we define $\mathbf{H}^{T}_{\textrm{BR}} \diamond  \mathbf{H}_{\textrm{RU}}$ as an effective cascaded channel $\mathbf{H}_{\textrm{eff}}$, which is a goal of channel estimation for RIS-aided MIMO systems.

With (\ref{HBR}), (\ref{HRU}), and Property 2, $\bm{\mathcal{Y}}$ can be fully unfolded as
\begin{equation}
    \begin{split}
    \bm{\mathcal{Y}}&=\frac{P_{\textrm{Tx}}}{N_{\textrm{B}}} \left(\mathbf{F}^{T} \otimes \mathbf{C}^{H}\right) \left( \mathbf{A}(\bm{\theta}_{\textrm{BR}})^{*} \otimes \mathbf{A}(\bm{\phi}_{\textrm{RU}})  \right)\\ & \left(\textrm{diag}(\bm{\rho}_{\textrm{BR}}) \otimes \textrm{diag}(\bm{\rho}_{\textrm{RU}}) \right) \left( \mathbf{A}(\bm{\phi}_{\textrm{BR}})^{T} \diamond \mathbf{A}(\bm{\theta}_{\textrm{RU}})^{H} \right) \mathbf{W} + \bm{\mathcal{V}}.
    \end{split}
\end{equation}
To simplify $\mathbf{A}(\bm{\phi}_{\textrm{BR}})^{T} \diamond \mathbf{A}(\bm{\theta}_{\textrm{RU}})^{H} \in \mathbb{C}^{L_{\textrm{BR}}L_{\textrm{RU}} \times M_{\textrm{R}}}$, $\bm{\varphi}$ is defined as
\begin{equation}
    \begin{split}
    \bm{\varphi}= \{ \varphi_{i,j}&: \cos^{-1} (\cos \theta^{j}_{\textrm{RU}} - \cos \phi^{i}_{\textrm{BR}} ), \\ & i=1,\ldots,L_{\textrm{BR}},j=1,\ldots,L_{\textrm{RU}}\}.
    \end{split}
\end{equation}
Then, $\mathbf{A}(\bm{\phi}_{\textrm{BR}})^{T} \diamond \mathbf{A}(\bm{\theta}_{\textrm{RU}})^{H}$ can be rewritten as
\begin{equation}
    \begin{split}
    &\mathbf{A}(\bm{\phi}_{\textrm{BR}})^{T} \diamond \mathbf{A}(\bm{\theta}_{\textrm{RU}})^{H}=\mathbf{A}(\bm{\varphi})^{H}\\&=\left[\mathbf{a}(\varphi_{1,1}),\ldots,\mathbf{a}(\varphi_{1,L_{\textrm{RU}}}),\ldots,\mathbf{a}(\varphi_{L_{\textrm{BR}},1}),\ldots,\mathbf{a}(\varphi_{L_{\textrm{BR}},L_{\textrm{RU}}}) \right]^{H}.
    \end{split}
\end{equation}

To evaluate the efficiency of the beam training by quantifying the quality of the pilot signals received during the beam training, we define an efficiency of the non location-aware beam training, $\textrm{eBT}_{\textrm{NLA}}$ as follows.
\begin{equation}\label{eSNR_NLA}
    \textrm{eBT}_{\textrm{NLA}}= \frac{\lVert \frac{P_{\textrm{Tx}}}{N_{\textrm{B}}} \left(\mathbf{F}^{T} \otimes \mathbf{C}^{H}\right) \left( \mathbf{H}^{T}_{\textrm{BR}} \diamond  \mathbf{H}_{\textrm{RU}} \right) \mathbf{W} \rVert^{2}_{\textrm{F}}}{\lVert \bm{\mathcal{V}} \rVert^{2}_{\textrm{F}}}.
\end{equation}
The beam training efficiency defined in (\ref{eSNR_NLA}) is a ratio between the pilot signal part and the noise part of $\bm{\mathcal{Y}}$. The beam training efficiency is a factor that significantly affects the performance of the channel estimation.

\subsection{Location-aware Beam Training}\label{LBT}
A frame structure for location-aware beam training is depicted in Fig.~\ref{Fig1_v2}.
Since $\theta^{1}_{\textrm{BR}}$ can be calculated by location of the BS and the RIS, the beam training procedure can be simplified by setting a precoding vector $\mathbf{f}$ as $\mathbf{a}(\theta^{1}_{\textrm{BR}})/\sqrt{M_{\textrm{B}}}$ during the beam training.
When the location information of BS and RIS are given, there are $P_{\textrm{U}}$ training symbols per frame, and the total number of training symbols equals to $BP_{\textrm{U}}$. 
Note that one RF chain of the BS is used during the beam training, and the UE still perform the exhaustive beam search as in non location-aware beam training.

When the location information is given, a received pilot signal at the $b$-th frame which uses the $i$-th combining matrix, $\mathbf{J}^{i}_{b}$ can be given by
\begin{equation}\label{rxsignal_loc}
    \mathbf{J}^{i}_{b}=\mathbf{C}_{i}^{H} \mathbf{H}_{\textrm{RU}} \bm{\Omega}_{b} \mathbf{H}_{\textrm{BR}} \mathbf{f} \mathbf{s} + \mathbf{N}^{i}_{b} \in \mathbb{C}^{N_{\textrm{U}} \times D},
\end{equation}
where $\mathbf{s} \in \mathbb{C}^{1 \times D}$ is the pilot signal that satisfies $\mathbf{s}\mathbf{s}^{H}/D=P_{\textrm{Tx}}$.
$\mathbf{N}^{i}_{b} \in \mathbb{C}^{N_{\textrm{U}} \times D}$ is a noise matrix whose columns follow $\mathcal{CN}(\mathbf{0}_{N_{\textrm{U}}},\sigma^{2} \mathbf{I}_{N_{\textrm{U}}})$. 

After receiving $\mathbf{J}^{i}_{b}$, $\mathbf{s}^{H}$ are multiplied to $\mathbf{J}^{i}_{b}$ to filter the noise, and the filtered signal $\mathbf{j}^{i}_{b}$ can be given by    
\begin{equation}\label{rxsignal2_loc}
    \mathbf{j}^{i}_{b}=\frac{\mathbf{J}^{i}_{b}\mathbf{s}^{H}}{D}=P_{\textrm{Tx}} \mathbf{C}_{j}^{H} \mathbf{H}_{\textrm{RU}} \bm{\Omega}_{b} \mathbf{H}_{\textrm{BR}} \mathbf{f} + \frac{\mathbf{N}^{i}_{b}\mathbf{s}^{H}}{D} \in \mathbb{C}^{N_{\textrm{U}} \times 1}.
\end{equation}
Then, a total $P_{\textrm{U}}$ filtered signals received at the $b$-th frame is organized as
\begin{equation}\label{merged_loc}
    \mathbf{j}_{b}=\left[(\mathbf{j}^{1}_{b})^{T}, (\mathbf{j}^{2}_{b})^{T}, \ldots ,(\mathbf{j}^{P_{\textrm{U}}}_{b})^{T} \right]^{T} \in \mathbb{C}^{M_{\textrm{U}} \times 1},
\end{equation}
where $\mathbf{j}_{b}$ is a compilation of all filtered signals at the $b$-th frame. $\mathbf{j}_{b}$ can also be represented as
\begin{equation}\label{merged2_loc}
    \mathbf{j}_{b}=P_{\textrm{Tx}} \mathbf{C}^{H} \mathbf{H}_{\textrm{RU}} \bm{\Omega}_{b} \mathbf{H}_{\textrm{BR}} \mathbf{f} + \mathbf{u}_{b},
\end{equation}
where $\mathbf{u}_{b} \in \mathbb{C}^{M_{\textrm{U}} \times 1}$ is a vector that represents the remaining noise and $\mathbf{u}_{b} \sim \mathcal{CN}\left( \mathbf{0}_{M_{\textrm{U}}},(\sigma^{2}P_{\textrm{Tx}}/D)\mathbf{I}_{M_{\textrm{U}}} \right)$.

To exploit the received pilot signals from all directions, $\mathbf{j}_{b}$ for $b=1,\ldots,B$ are organized in the same manner that is introduced in Section~\ref{NLBT}.
Since organizing $\mathbf{j}_{b}$ for $b=1,\ldots,B$ is not much different from the organization in Section~\ref{NLBT}, the detailed step of organizing $\mathbf{j}_{b}$ for $b=1,\ldots,B$ is omitted.
An organized matrix $\bm{\mathcal{J}}$ is constructed by stacking $\mathbf{j}_{b}$ for $b=1,\ldots,B$ as  
\begin{equation}
    \begin{split}
    \bm{\mathcal{J}}&=\left[\mathbf{j}_{1},\ldots,\mathbf{j}_{B} \right]\\&=P_{\textrm{Tx}} \left(\mathbf{f}^{T} \mathbf{H}^{T}_{\textrm{BR}} \diamond \mathbf{C}^{H} \mathbf{H}_{\textrm{RU}}\right) \mathbf{W}  + \mathbf{U} \\
    &= P_{\textrm{Tx}} \left(\mathbf{f}^{T} \otimes \mathbf{C}^{H} \right) \left(\mathbf{H}^{T}_{\textrm{BR}} \diamond \mathbf{H}_{\textrm{RU}} \right)\mathbf{W} + \mathbf{U} \in \mathbb{C}^{M_{\textrm{U}} \times B},
    \end{split}
\end{equation}
where $\mathbf{U}=[\mathbf{u}_{1},\ldots,\mathbf{u}_{B}] \in \mathbb{C}^{M_{\textrm{U}} \times B}$. $\bm{\mathcal{J}}$ can be fully unfolded as
\begin{equation}
    \begin{split}
    \bm{\mathcal{J}}=&P_{\textrm{Tx}} \left(\mathbf{f}^{T} \otimes \mathbf{C}^{H}\right) \left( \mathbf{A}(\bm{\theta}_{\textrm{BR}})^{*} \otimes \mathbf{A}(\bm{\phi}_{\textrm{RU}})  \right)\\ & \left(\textrm{diag}(\bm{\rho}_{\textrm{BR}}) \otimes \textrm{diag}(\bm{\rho}_{\textrm{RU}}) \right) \mathbf{A}(\bm{\varphi})^{H} \mathbf{W} + \mathbf{U}.
    \end{split}
\end{equation}

As in~(\ref{eSNR_NLA}), we define the efficiency of the location-aware beam training, $\textrm{eBT}_{\textrm{LA}}$ as follows.
\begin{equation}\label{eSNR_LA}
    \textrm{eBT}_{\textrm{LA}}= \frac{\lVert P_{\textrm{Tx}} \left(\mathbf{f}^{T} \otimes \mathbf{C}^{H} \right) \left(\mathbf{H}^{T}_{\textrm{BR}} \diamond \mathbf{H}_{\textrm{RU}} \right)\mathbf{W} \rVert^{2}_{\textrm{F}}}{\lVert \mathbf{U} \rVert^{2}_{\textrm{F}}}.
\end{equation}
The beam training efficiency defined in (\ref{eSNR_LA}) is a ratio between the pilot signal part and the noise part of $\bm{\mathcal{J}}$.

\begin{figure*}
    \begin{center}
    \captionsetup[subfigure]{justification=centering}
    \begin{subfigure}[t]{0.82\columnwidth}
        \includegraphics[width=\columnwidth,center]{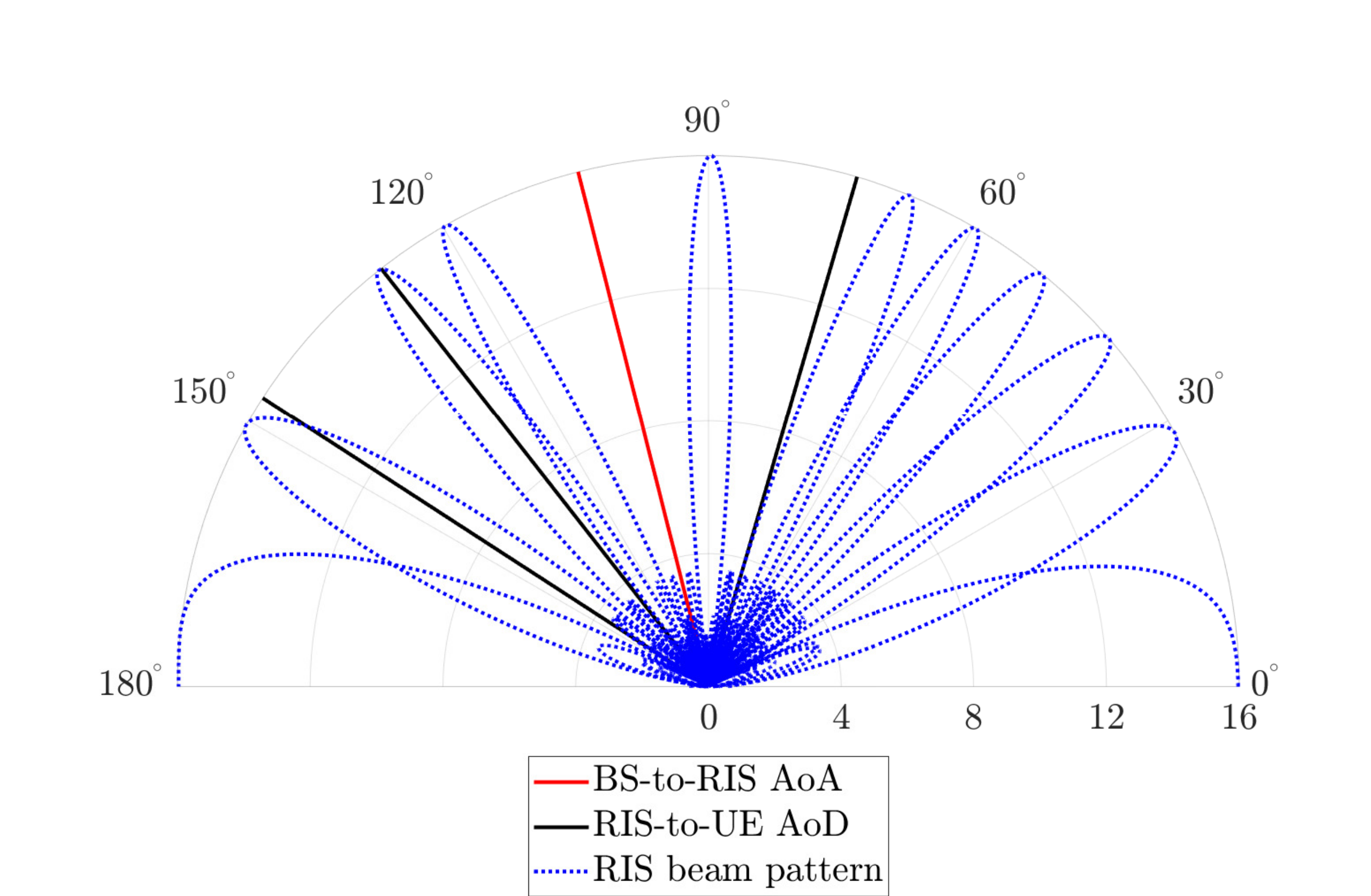}
        \caption{Without RIS beamwidth adaptation}\label{2a}
    \end{subfigure}
    \begin{subfigure}[t]{0.82\columnwidth}
        \includegraphics[width=\columnwidth,center]{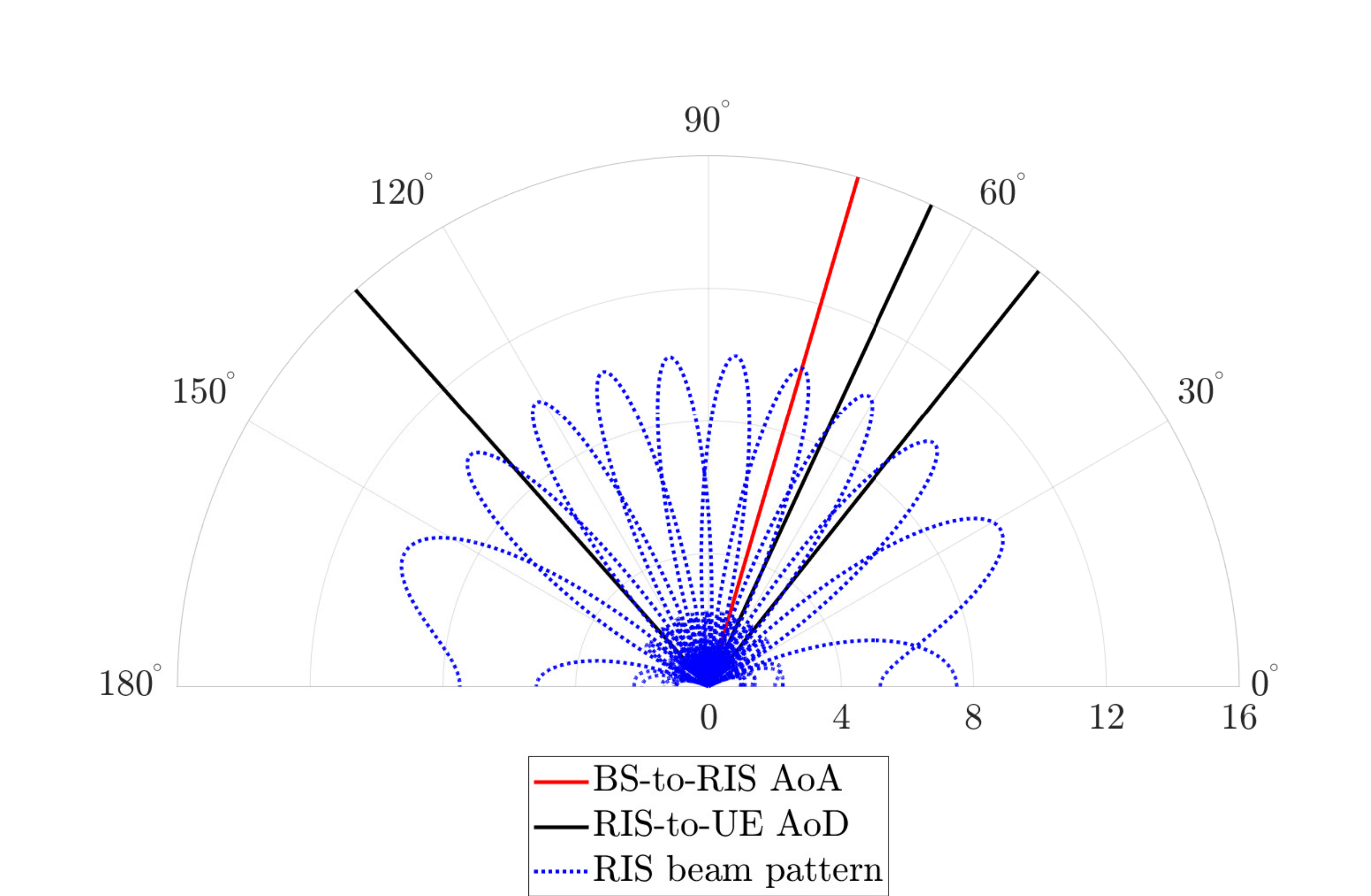}
        \caption{With RIS beamwidth adaptation}\label{2b}
    \end{subfigure}
    \caption{Reflecting radiation patterns created by RIS during the beam training. $L_{\textrm{BR}}=1$, $L_{\textrm{RU}}=3$, $M_{\textrm{R}}=16$, and $B=10$. For successful multipath signal reception, every RIS-to-UE AoD has to be captured within one of reflect beams.} 
    \label{all}
    \end{center}
\end{figure*}

\section{RIS Beamwidth Adaptation for Robust Low-Overhead Channel Estimation}\label{RISBWA}
During the beam training, the RIS must deliver the signal transmitted from the BS to the UE by sweeping reflect beam over entire angular domain.
If not, there is a possibility that the UE does not receive the signal transmitted from the BS, which leads to the channel estimation failure. 
A reflecting radiation pattern of RIS, $\Gamma(\theta,\bar{\phi},\bm{\omega})$ can be given by 
\begin{equation}
    \Gamma(\theta,\bar{\phi},\bm{\omega})=\left\vert \mathbf{a}(\theta)^{H} \textrm{diag}(\bm{\omega}) \mathbf{a}(\bar{\phi}) \right\vert,\;\textrm{for}\;0^{\circ} < \bar{\phi} < 180^{\circ}, 
\end{equation}
where $\theta$, $\bar{\phi}$, and $\bm{\omega}$ respectively denote the BS-to-RIS AoA, the steering direction of the reflect beam, and the RIS control vector. 
The received signal is boosted if the RIS-to-UE AoD is on the mainlobe of the reflect beam. Otherwise, the signal is suppressed or nulled.   

One of general ways to perform reflect beamforming is to use a discrete Fourier transform (DFT) matrix~\cite{5707050}. $N \times N$ DFT matrix $\bm{\Psi}_{N}$ can be given by
\begin{equation}
    \bm{\Psi}_{N}= 
    \begin{bmatrix}
    1 & 1 & \cdots & 1 \\
    1 & e^{j \frac{2\pi}{N}} & \cdots & e^{j \frac{2\pi (N-1)}{N}} \\
    1 & e^{j \frac{4\pi}{N}} & \cdots & e^{j \frac{4\pi (N-1)}{N}} \\
    \vdots & \vdots & \cdots & \vdots \\
    1 & e^{j 2\pi} & \cdots & e^{j 2\pi(N-1)}
    \end{bmatrix}
    \in \mathbb{C}^{N \times N}.
\end{equation}
For mainlobes of reflect beams to cover the entire angular domain during the beam training, $\mathbf{W}$ should be equal to $\bm{\Psi}_{M_{\textrm{R}}}$ so that $B=M_{\textrm{R}}$.
However, considering the total number of training symbol is proportional to $B$, $B$ should be reduced in order to prevent the beam training overhead from getting excessively large.
As in~\cite{9354904}, a few columns from $\bm{\Psi}_{M_{\textrm{R}}}$ can be selected to form $\mathbf{W}$, but lack of reflect beams can cause an erroneous multipath signal reception.
Fig.~\ref{2a} shows a case of the erroneous multipath signal reception when none of RIS-to-UE signal paths falls onto the mainlobes of reflect beams. 
In this case, the UE cannot receive multipath components during the beam training, and the channel estimation fails since signals from all paths are required for perfect channel estimation.  

To address this issue, a RIS beamwidth adaptation is proposed to make multipath signal reception robust when $B<M_{\textrm{R}}$.
The width of reflect beams can be widened by deactivating the part of RIS, and $\mathbf{W}$ that contains $B$ widened reflect beams can be given by
\begin{equation}
    \mathbf{W}=
    \begin{bmatrix}
    \bm{\Psi}_{B} \\ \mathbf{O}_{M_{\textrm{R}}-B,B}
    \end{bmatrix} \in \mathbb{C}^{M_\textrm{R} \times B},
\end{equation}
where $\mathbf{O}_{M,N}$ denotes a $M \times N$ zero matrix. Note that only $B$ antennas are activated, and other antennas are not. 
Fig.~\ref{2b} shows beams that are widened by RIS beamwidth adaptation when $M_{\textrm{R}}=16$ and $B=10$. 
In Fig.~\ref{2b}, mainlobes of $B$ beams cover the entire angular domain so that every RIS-to-UE signal path is captured within one of $B$ beams. 
However, the deactivating the part of the RIS drops the power of the received signal and may degrade an accuracy of the channel estimation.

For the following sections of this paper, notations $\mathbf{W}_{0}$ and $\mathbf{W}_{1}$ are defined to distinguish whether the RIS beamwidth adaptation is used or not.  
$\mathbf{W}_{0}$ stands for the set of the RIS control vectors when the RIS beamwidth adaptation is not employed, so that $\mathbf{W}_{0}=\bm{\Psi}_{M_{\textrm{R}}}$. 
$\mathbf{W}_{1}$ stands for the set of the RIS control vectors when the RIS beamwidth adaptation is employed. Also, $B$ is considered to be always less than $M_{\textrm{R}}$.

\section{Low-overhead Channel Estimation via Atomic Norm Minimization}\label{alg}
In this section, we introduce the non location-aware channel estimation and the location-aware channel estimation based on ANM.
If the location information of BS and RIS are not given, the channel can be estimated via 1D ANM.
When the location information of BS and RIS is given, there are two options to estimate the channel depending on whether the RIS beamwidth adaptation is used or not. Two options are:
\begin{itemize}
    \item Option 1: Perform exhaustive beam search at the RIS. In this case, 2D ANM is employed for the channel estimation.
    \item Option 2: Perform RIS beamwidth adaptation. If so, the beam training overhead can be further reduced. In this case, 3D ANM is employed for the channel estimation.
\end{itemize}
To ease the understanding of the atomic norm, an explanation of atom, atomic set, and Toeplitz matrix is given in the following subsection.

\subsection{Atom, Atomic set, and Toeplitz Matrix}
If the signal can be represented as the linear combination of basic components, the atomic norm of the signal can be defined. 
Here, the basic component is referred to as an atom, and a set of atoms is referred to as an atomic set.
The atomic norm has been studied in the seminal works, especially for a case when the atom can be represented as the steering vector of the ULA~\cite{6576276,7313018}.
In this case, the atomic set $\mathcal{A}_{\textrm{1D}}$ is generally defined as 
\begin{equation}
    \begin{split}
    \mathcal{A}_{\textrm{1D}}=&\big\{\mathbf{a}(\theta)\mathbf{b}^{T} \in \mathbb{C}^{M \times P}: 0^{\circ} < \theta < 180^{\circ}, \\ & \quad \mathbf{a}(\theta) \in \mathbb{C}^{M \times 1}, \mathbf{b} \in \mathbb{C}^{P \times 1}, \lVert \mathbf{b} \rVert_{2}=1  \big\},        
    \end{split}
\end{equation}
where $\mathbf{a}(\theta)\mathbf{b}^{T}$ denotes the atom. A case when $P=1$ is referred to as single measurement vector (SMV) case~\cite{6576276}, and a case when $P>1$ is referred to as multiple measurement vector (MMV) case~\cite{7313018}.     
An atomic norm that is derived based on $\mathcal{A}_{\textrm{1D}}$ is referred to as 1D atomic norm. 
To represent the 1D atomic norm with SDP, a Hermitian Toeplitz matrix is used, where a $M \times M$ Hermitian Toeplitz matrix $\mathbf{T}$ can be given by
\begin{equation}\label{Toepu}
\mathbf{T}=
\begin{bmatrix}
a_1 & a^{*}_2 & \ldots & a^{*}_{M-1} & a^{*}_M \\
a_2 & a_1 & a^{*}_2 & \ldots & a^{*}_{M-1} \\
\vdots & a_2 & \ddots & \ddots & \vdots \\
a_{M-1} & \vdots & \ddots & \ddots & a_2  \\
a_M & a_{M-1} & \ldots & a_2 & a_1
\end{bmatrix} \in \mathbb{C}^{M \times M}.
\end{equation}

However, there are cases that the atom is modeled as a result of the Kronecker product of multiple steering vectors~\cite{7451201}. In this paper, we refer the atomic norm for such cases as a \textit{multi-dimensional atomic norm}.
Most well-known cases that employ multi-dimensional atomic norm are joint azimuth/elevation estimation~\cite{7952721} and MIMO channel estimation~\cite{8432470}.
An atomic set comprised of $N$-dimensional ($N$D) atoms can be defined as
\begin{equation}\label{NDatom}
    \begin{split}
    \mathcal{A}_{N\textrm{D}}=\big\{ \left(\mathbf{a}_{1}(\theta_{1})^{*} \otimes \ldots \otimes \mathbf{a}_{N}(\theta_{N})\right) \mathbf{b}^{T} \in \mathbb{C}^{(\prod_{n=1}^{N}M_{n}) \times P}: \\
    \mathbf{b} \in \mathbb{C}^{P \times 1}, \lVert \mathbf{b} \rVert_{2}=1, 0^{\circ} < \theta_{n} < 180^{\circ},\; \textrm{for} \; n=1,\ldots,N \big\},
    \end{split}
\end{equation}
where $\left(\mathbf{a}_{1}(\theta_{1})^{*} \otimes \ldots \otimes \mathbf{a}_{N}(\theta_{N})\right) \mathbf{b}^{T}$ is the $N$D atom, and $\mathbf{a}_{n}(\theta_{n}) \in \mathbb{C}^{M_{n} \times 1}$ denotes the $n$-th steering vector that composes the $N$D atom.
The SDP representation the $N$D atomic norm requires a $N$-level Toeplitz matrix $\mathbf{T}_{N\textrm{D}}$, which can be given by follows~\cite{7451201}:
\begin{equation}\label{NDToep}
    \begin{split}
    \mathbf{T}_{N\textrm{D}} &= \sum_{r=1}^{R} d_{r} \mathbf{a}_{N\textrm{D},r} \mathbf{a}^{H}_{N\textrm{D},r} \\ &= \mathbf{A}_{N\textrm{D}} \mathbf{D} \mathbf{A}^{H}_{N\textrm{D}} \in \mathbb{C}^{(\prod_{n=1}^{N}M_{n}) \times (\prod_{n=1}^{N}M_{n})}. 
    \end{split}
\end{equation}
$\mathbf{a}_{N\textrm{D},r}$ denotes the $r$-th $N$D atom, which can be given by $\left(\mathbf{a}_{1}(\theta_{1,r})^{*} \otimes \ldots \otimes \mathbf{a}_{N}(\theta_{N,r})\right) \mathbf{b}_{r}^{T}$.
Here, $\theta_{n,r}$ denotes the $r$-th angle of the $n$-th steering vector. 
$d_{r}$ denotes the complex coefficient of the $r$-th $N$D atom.
$\mathbf{A}_{N\textrm{D}}=\left[ \mathbf{a}_{N\textrm{D},1},\ldots,\mathbf{a}_{N\textrm{D},R} \right]$, and $\mathbf{D}=\textrm{diag}(\mathbf{d})$, where $\mathbf{d}=\left[d_{1},\ldots,d_{R} \right]^{T} \in \mathbb{C}^{R \times 1}$. 

\subsection{Non Location-aware Channel Estimation via 1D ANM}\label{1DANM}
When performing the RIS beamwidth adaptation without using the location information, the effective cascaded channel can be estimated via 1D ANM.
To simplify equations and notations, we define $\mathbf{G}_{\textrm{1D}}$ and $\mathbf{Z}_{\textrm{1D}}$ as follows.
\begin{equation}
    \begin{split}
    \mathbf{G}_{\textrm{1D}}=\left\{ \left(\mathbf{F}^{T} \otimes \mathbf{C}^{H}\right)^{-1} \bm{\mathcal{Y}} \right\}^{H}=\mathbf{W}_{1}^{H} \mathbf{Z}_{\textrm{1D}} + \mathbf{E}_{\textrm{1D}},  
    \end{split}
\end{equation}
\begin{equation}
    \begin{split}
    &\mathbf{Z}_{\textrm{1D}}=\frac{P_{\textrm{Tx}}}{N_{\textrm{B}}} \mathbf{H}_{\textrm{eff}}^{H}=\frac{P_{\textrm{Tx}}}{N_{\textrm{B}}} \left( \mathbf{H}^{T}_{\textrm{BR}} \diamond  \mathbf{H}_{\textrm{RU}} \right)^{H}=\\ &\frac{P_{\textrm{Tx}}}{N_{\textrm{B}}} \mathbf{A}(\bm{\varphi}) \left(\textrm{diag}(\bm{\rho}_{\textrm{BR}}) \otimes \textrm{diag}(\bm{\rho}_{\textrm{RU}}) \right)^{H} \left( \mathbf{A}(\bm{\theta}_{\textrm{BR}})^{*} \otimes \mathbf{A}(\bm{\phi}_{\textrm{RU}})  \right)^{H}.
    \end{split}
\end{equation}
Here, $\mathbf{E}_{\textrm{1D}}=\{ \left(\mathbf{F}^{T} \otimes \mathbf{C}^{H}\right)^{-1} \bm{\mathcal{V}} \}^{H}$, and each column of $\mathbf{E}_{\textrm{1D}}$ follows $\mathcal{CN}\left(\mathbf{0}_{M_{\textrm{B}}M_{\textrm{U}}},(\sigma^{2}P_{\textrm{Tx}}/DN_{\textrm{B}})\mathbf{I}_{M_{\textrm{B}}M_{\textrm{U}}}\right)$. 
To derive the 1D atomic norm of $\mathbf{Z}_{\textrm{1D}}$, an atomic set $\mathcal{A}_{\textrm{1D}}$ is defined as follows.
\begin{equation}\label{atom}
    \begin{split}
    \mathcal{A}_{\textrm{1D}}=&\big\{\mathbf{a}(\theta)\mathbf{b}^{T} \in \mathbb{C}^{M_{\textrm{R}} \times M_{\textrm{B}}M_{\textrm{U}}}: 0^{\circ} < \theta < 180^{\circ}, \\ & \mathbf{a}(\theta) \in \mathbb{C}^{M_{\textrm{R}} \times 1}, \mathbf{b} \in \mathbb{C}^{M_{\textrm{B}}M_{\textrm{U}} \times 1}, \lVert \mathbf{b} \rVert_{2}=1  \big\},        
    \end{split}
\end{equation}
where $\mathbf{a}(\theta)\mathbf{b}^{T}$ is an atom of $\mathbf{Z}_{\textrm{1D}}$ since $\mathbf{Z}_{\textrm{1D}}$ can be represented as a linear combination of atoms.
Since $M_{\textrm{B}}M_{\textrm{U}} > 1$, the 1D atomic norm of $\mathbf{Z}_{\textrm{1D}}$ is considered as the MMV case.
The 1D atomic norm of $\mathbf{Z}_{\textrm{1D}}$ for the MMV case, $\lVert \mathbf{Z}_{\textrm{1D}} \rVert_{\mathcal{A}_{\textrm{1D}}}$ can be represented by following SDP~\cite{9016105}: 
\begin{equation}
    \begin{split}   
    \lVert \mathbf{Z}_{\textrm{1D}} \rVert_{\mathcal{A}_{\textrm{1D}}}=&\min_{\mathbf{T},\mathbf{P}_{\textrm{1D}}} \;  \frac{1}{2M_{\textrm{R}}} \textrm{Tr}(\mathbf{T})+ \frac{1}{2} \textrm{Tr}(\mathbf{P}_{\textrm{1D}}) \\ &\; \textrm{s.t.}
    \begin{bmatrix}
    \mathbf{T} & \mathbf{Z}_{\textrm{1D}} \\
    \mathbf{Z}_{\textrm{1D}}^{H} & \mathbf{P}_{\textrm{1D}}
    \end{bmatrix} \succeq 0.
    \end{split}
\end{equation}
With the ANM denoising theorem studied in~\cite{7313018}, an equation that estimates $\mathbf{Z}_{\textrm{1D}}$ from $\mathbf{G}_{\textrm{1D}}$ can be given by
\begin{equation}\label{final}   
    \hat{\mathbf{Z}}_{\textrm{1D}}=\argmin_{\bar{\mathbf{Z}}_{\textrm{1D}}} \;  \tau_{\textrm{1D}} \lVert \bar{\mathbf{Z}}_{\textrm{1D}} \rVert_{\mathcal{A}_{\textrm{1D}}} + \frac{\lVert \mathbf{G}_{\textrm{1D}}-\mathbf{W}_{1}^{H}\bar{\mathbf{Z}}_{\textrm{1D}}\rVert_{\textrm{F}}^{2}}{2},
\end{equation}
where $\hat{\mathbf{Z}}_{\textrm{1D}}$ and $\bar{\mathbf{Z}}_{\textrm{1D}}$ respectively denote the estimate of $\mathbf{Z}_{\textrm{1D}}$ and the variable for estimation of $\mathbf{Z}_{\textrm{1D}}$. 
Considering the covariance of each element of $\mathbf{E}_{\textrm{1D}}$, a regularization parameter $\tau_{\textrm{1D}}$ is set as in~\cite{7313018}: 
\begin{equation}\label{tau}   
    \begin{split}
    \tau_{\textrm{1D}}=&\frac{\sigma\sqrt{P_{\textrm{Tx}}}}{\sqrt{DN_{\textrm{B}}}}\Big(1+\frac{1}{\log M_{\textrm{R}}} \Big)^{\frac{1}{2}} \Big( M_{\textrm{B}}M_{\textrm{U}} + \log(\alpha_{\textrm{1D}} M_{\textrm{B}}M_{\textrm{U}}) +  \\  &\sqrt{2M_{\textrm{B}}M_{\textrm{U}} \log(\alpha_{\textrm{1D}} M_{\textrm{B}}M_{\textrm{U}})} + \sqrt{\frac{\pi M_{\textrm{B}}M_{\textrm{U}}}{2}} +1  \Big)^{\frac{1}{2}},
    \end{split}
\end{equation}
where $\alpha_{\textrm{1D}}=8\pi M_{\textrm{R}} \log M_{\textrm{R}}$.
(\ref{final}) can be fully unfolded as
\begin{equation}\label{final2}   
    \begin{split}
    &\{\hat{\mathbf{T}},\hat{\mathbf{P}}_{\textrm{1D}},\hat{\mathbf{Z}}_{\textrm{1D}} \}=\\ & \argmin_{\mathbf{T},\mathbf{P}_{\textrm{1D}},\bar{\mathbf{Z}}_{\textrm{1D}}} \;  \frac{\tau_{\textrm{1D}}\textrm{Tr}(\mathbf{T})}{2M_{\textrm{R}}} + \frac{\tau_{\textrm{1D}}\textrm{Tr}(\mathbf{P}_{\textrm{1D}})}{2} + \frac{\lVert \mathbf{G}_{\textrm{1D}}-\mathbf{W}_{1}^{H}\bar{\mathbf{Z}}_{\textrm{1D}}\rVert_{\textrm{F}}^{2}}{2} \\ &\; \textrm{s.t.}
    \begin{bmatrix}
    \mathbf{T} & \bar{\mathbf{Z}}_{\textrm{1D}} \\
    \bar{\mathbf{Z}}_{\textrm{1D}}^{H} & \mathbf{P}_{\textrm{1D}}
    \end{bmatrix} \succeq 0. 
    \end{split}
\end{equation}
Finally, $\mathbf{H}_{\textrm{eff}}$ can be approximated as $(N_{\textrm{B}}/P_{\textrm{Tx}})\hat{\mathbf{Z}}_{\textrm{1D}}^{H}$. 

A computational complexity of (\ref{final2}) can be derived by a theorem in~\cite{SDPcomp} after transcribing (\ref{final2}) into a general SDP form. 
If an influence of a stopping criteria for the optimization is not considered, $\mathcal{O}\left( (M_{\textrm{B}}M_{\textrm{U}}+M_{\textrm{R}})^{0.5} \right)$ iterations are required, where each iteration costs $\mathcal{O}\left( (M_{\textrm{B}}M_{\textrm{U}}+M_{\textrm{R}})^{3} \right)$.
Thus, the computational complexity of the non location-aware channel estimation via 1D ANM is $\mathcal{O}\left( (M_{\textrm{B}}M_{\textrm{U}}+M_{\textrm{R}})^{3.5} \right)$.

\subsection{Location-aware Channel Estimation via 2D ANM}
If the location information is given and the RIS performs the exhaustive search, the effective cascaded channel can be estimated via 2D ANM.
To simplify equations and notations, we define $\mathbf{G}_{\textrm{2D}}$ and $\mathbf{Z}_{\textrm{2D}}$ as follows.
\begin{equation}
    \begin{split}
    \mathbf{G}_{\textrm{2D}}=\bm{\mathcal{J}} \mathbf{W}_{0}^{-1}= \left(\mathbf{f}^{T} \otimes \mathbf{C}^{H}\right) \mathbf{Z}_{\textrm{2D}} + \mathbf{E}_{\textrm{2D}} \in \mathbf{C}^{M_{\textrm{U}} \times M_{\textrm{R}}},  
    \end{split}
\end{equation}
\begin{equation}
    \begin{split}
    &\mathbf{Z}_{\textrm{2D}}=P_{\textrm{Tx}}\mathbf{H}_{\textrm{eff}}=\\
    &P_{\textrm{Tx}}\left( \mathbf{A}(\bm{\theta}_{\textrm{BR}})^{*} \otimes \mathbf{A}(\bm{\phi}_{\textrm{RU}})  \right) \left(\textrm{diag}(\bm{\rho}_{\textrm{BR}}) \otimes \textrm{diag}(\bm{\rho}_{\textrm{RU}}) \right) \mathbf{A}(\bm{\varphi})^{H}.
    \end{split}
\end{equation}
Here, $\mathbf{E}_{\textrm{2D}}= \mathbf{U}\mathbf{W}_{0}^{-1}$, and each column of $\mathbf{E}_{\textrm{2D}}$ follows $\mathcal{CN}\left( \mathbf{0}_{M_{\textrm{U}}},(\sigma^{2}P_{\textrm{Tx}}/DM_{\textrm{R}})\mathbf{I}_{M_{\textrm{U}}} \right)$. 
To estimate $\mathbf{Z}_{\textrm{2D}}$ using $\mathbf{G}_{\textrm{2D}}$, the atomic norm and the atomic set need to be determined.
An atomic norm of $\mathbf{Z}_{\textrm{2D}}$, $\mathcal{A}_{\textrm{2D}}$ can be represented as follows.
\begin{equation}
\begin{split}
    \mathcal{A}_{\textrm{2D}}=&\left\{ \left( \mathbf{a}_{1}(\theta_{1})^{*} \otimes \mathbf{a}_{2}(\theta_{2}) \right) \mathbf{b}^{T} \in \mathbb{C}^{M_{\textrm{B}}M_{\textrm{U}} \times M_{\textrm{R}}}: \right. \\ & \left. \mathbf{b} \in \mathbb{C}^{M_{\textrm{R}} \times 1},\; \lVert \mathbf{b} \rVert_{2}=1,\; 0^{\circ} < \theta_{1},\theta_{2} < 180^{\circ} \right\},        
\end{split}
\end{equation}
where $\left(\mathbf{a}_{1}(\theta_{1})^{*} \otimes \mathbf{a}_{2}(\theta_{2}) \right) \mathbf{b}^{T}$ is an atom for $\mathbf{Z}_{\textrm{2D}}$, and $\mathbf{Z}_{\textrm{2D}}$ can be represented as a linear combination of these atoms.
If the atomic set of $\mathbf{Z}_{\textrm{2D}}$ is $\mathcal{A}_{\textrm{2D}}$, the atomic norm for $\mathbf{Z}_{\textrm{2D}}$, $\lVert \mathbf{Z}_{\textrm{2D}} \rVert_{\mathcal{A}_{\textrm{2D}}}$ can be given by following SDP~\cite{7451201}:
\begin{equation}
    \begin{split}   
    \lVert \mathbf{Z}_{\textrm{2D}} \rVert_{\mathcal{A}_{\textrm{2D}}}=&\min_{\mathbf{T}_{\textrm{2D}},\mathbf{P}} \;  \frac{1}{2M_{\textrm{B}}M_{\textrm{U}}} \textrm{Tr}(\mathbf{T}_{\textrm{2D}})+ \frac{1}{2} \textrm{Tr}(\mathbf{P}) \\ &\; \textrm{s.t.}
    \begin{bmatrix}
    \mathbf{T}_{\textrm{2D}} & \mathbf{Z}_{\textrm{2D}} \\
    \mathbf{Z}_{\textrm{2D}}^{H} & \mathbf{P}
    \end{bmatrix} \succeq 0,
    \end{split}
\end{equation}
where $\mathbf{T}_{\textrm{2D}}$ denotes $2$-level Toeplitz matrix. 
The structure of $\mathbf{T}_{\textrm{2D}}$ can be represented by Hermitian Toeplitz matrices and Toeplitz matrices as follows.  
\begin{equation}\label{2Dstructure}
    \mathbf{T}_{\textrm{2D}}= 
    \begin{bmatrix}
    \mathbf{T}_{0} & \mathbf{T}_{1}^{H} & \cdots & \mathbf{T}_{M_{\textrm{B}}-1}^{H} & \mathbf{T}_{M_{\textrm{B}}}^{H} \\
    \mathbf{T}_{1} & \mathbf{T}_{0} & \mathbf{T}_{1}^{H} & \ddots & \mathbf{T}_{M_{\textrm{B}}-1}^{H} \\
    \vdots & \ddots & \ddots & \ddots & \vdots \\
    \mathbf{T}_{M_{\textrm{B}}-1} & \ddots & \mathbf{T}_{1} & \mathbf{T}_{0} & \mathbf{T}_{1}^{H} \\
    \mathbf{T}_{M_{\textrm{B}}} & \mathbf{T}_{M_{\textrm{B}}-1} & \cdots & \mathbf{T}_{1} & \mathbf{T}_{0}
    \end{bmatrix}.
\end{equation}
Here, $\mathbf{T}_{0}$ denotes a Hermitian Toeplitz matrix. For $i=1,\ldots,M_{\textrm{B}}$, $\mathbf{T}_{i}$ denotes a Toeplitz matrix.
The structure of the Toeplitz matrix is different from that of the Hermitian Toeplitz matrix, where the structure of $M_{\textrm{B}}$ Toeplitz matrices that compose $\mathbf{T}_{\textrm{2D}}$ can be given by
\begin{equation}\label{NormalToep}
\begin{split}
\mathbf{T}_{i}=
\begin{bmatrix}
a_1 & a_{-2} & \ldots & a_{-M_{\textrm{U}}+1} & a_{-M_{\textrm{U}}} \\
a_2 & a_1 & a_{-2} & \ldots & a_{-M_{\textrm{U}}+1} \\
\vdots & a_2 & \ddots & \ddots & \vdots \\
a_{M_{\textrm{U}}-1} & \vdots & \ddots & \ddots & a_{-2}  \\
a_{M_{\textrm{U}}} & a_{M_{\textrm{U}}-1} & \ldots & a_2 & a_1
\end{bmatrix},\\
\textrm{for}\; i=1,\ldots,M_{\textrm{B}}.
\end{split}
\end{equation}

To simplify the representation of following equations, we define $\mathbf{Q}$ as $\mathbf{f}^{T} \otimes \mathbf{C}^{H}$.
An equation that estimates $\mathbf{Z}_{\textrm{2D}}$ from $\mathbf{G}_{\textrm{2D}}$ can be given by
\begin{equation}\label{2Ddenoise}   
    \hat{\mathbf{Z}}_{\textrm{2D}}=\argmin_{\bar{\mathbf{Z}}_{\textrm{2D}}} \;  \tau_{\textrm{2D}} \lVert \bar{\mathbf{Z}}_{\textrm{2D}} \rVert_{\mathcal{A}_{\textrm{2D}}} + \frac{1}{2} \lVert \mathbf{G}_{\textrm{2D}}-\mathbf{Q}\bar{\mathbf{Z}}_{\textrm{2D}} \rVert_{\textrm{F}}^{2},
\end{equation}
where $\hat{\mathbf{Z}}_{\textrm{2D}}$ and $\bar{\mathbf{Z}}_{\textrm{2D}}$ respectively denote the estimate of $\mathbf{Z}_{\textrm{2D}}$ and the variable for estimation of $\mathbf{Z}_{\textrm{2D}}$. 
Considering the covariance of each element of $\mathbf{E}_{\textrm{2D}}$, a regularization parameter $\tau_{\textrm{2D}}$ is set as
\begin{equation}\label{tau2}   
    \begin{split}
    \tau_{\textrm{2D}}=&\frac{\sigma\sqrt{P_{\textrm{Tx}}}}{\sqrt{DM_{\textrm{R}}}}\Big(1+\frac{1}{\log M_{\textrm{B}}M_{\textrm{U}}} \Big)^{\frac{1}{2}} \Big( M_{\textrm{R}} + \log(\alpha_{\textrm{2D}} M_{\textrm{R}}) +  \\  &\sqrt{2M_{\textrm{R}} \log(\alpha_{\textrm{2D}} M_{\textrm{R}})} + \sqrt{\frac{\pi M_{\textrm{R}}}{2}} +1  \Big)^{\frac{1}{2}},
    \end{split}
\end{equation}
where $\alpha_{\textrm{2D}}=8\pi M_{\textrm{B}}M_{\textrm{U}} \log \left( M_{\textrm{B}}M_{\textrm{U}} \right)$.
(\ref{2Ddenoise}) can be fully unfolded as
\begin{equation}\label{2Dunfold} 
    \begin{split}
    &\{\hat{\mathbf{T}}_{\textrm{2D}},\hat{\mathbf{P}}_{\textrm{2D}},\hat{\mathbf{Z}}_{\textrm{2D}} \}=\\ & \argmin_{\mathbf{T}_{\textrm{2D}},\mathbf{P}_{\textrm{2D}},\bar{\mathbf{Z}}_{\textrm{2D}}} \;  \frac{\tau_{\textrm{2D}}\textrm{Tr}(\mathbf{T}_{\textrm{2D}})}{2M_{\textrm{B}}M_{\textrm{U}}} + \frac{\tau_{\textrm{2D}}\textrm{Tr}(\mathbf{P}_{\textrm{2D}})}{2} + \frac{\lVert \mathbf{G}_{\textrm{2D}}-\mathbf{Q}\bar{\mathbf{Z}}_{\textrm{2D}} \rVert_{\textrm{F}}^{2}}{2} \\ &\; \textrm{s.t.}
    \begin{bmatrix}
    \mathbf{T}_{\textrm{2D}} & \bar{\mathbf{Z}}_{\textrm{2D}} \\
    \bar{\mathbf{Z}}_{\textrm{2D}}^{H} & \mathbf{P}_{\textrm{2D}}
    \end{bmatrix} \succeq 0. 
    \end{split}
\end{equation}
After solving (\ref{2Dunfold}), $\mathbf{H}_{\textrm{eff}}$ can be approximated as $\hat{\mathbf{Z}}_{\textrm{2D}}/P_{\textrm{Tx}}$.

A computational complexity of (\ref{2Dunfold}) can be derived by a theorem in~\cite{SDPcomp} after transcribing (\ref{2Dunfold}) into a general SDP form. 
If an influence of a stopping criteria for the optimization is not considered, $\mathcal{O}\left( (M_{\textrm{B}}M_{\textrm{U}}+M_{\textrm{R}})^{0.5} \right)$ iterations are required, where each iteration costs $\mathcal{O}\left( (M_{\textrm{B}}M_{\textrm{U}}+M_{\textrm{R}})^{3} \right)$.
Thus, the computational complexity of the location-aware channel estimation via 2D ANM is $\mathcal{O}\left( (M_{\textrm{B}}M_{\textrm{U}}+M_{\textrm{R}})^{3.5} \right)$.

\subsection{Location-aware Channel Estimation via 3D ANM}
By performing the RIS beamwidth adaptation when the location information is given, the beam training overhead can be further reduced, and the effective cascaded channel can be estimated via 3D ANM. The 3D ANM-based low overhead channel estimation requires the vectorization of $\bm{\mathcal{J}}$, where the vectorized $\bm{\mathcal{J}}$, $\bm{\Upsilon}$ can be represented as 
\begin{equation}\label{3Dvec}
\begin{split}
\bm{\Upsilon} =& \textrm{vec}(\bm{\mathcal{J}})=P_{\textrm{Tx}}\left(\mathbf{W}_{1}^{T} \otimes \mathbf{f}^{T} \otimes \mathbf{C}^{H}\right)\textrm{vec}(\mathbf{H}_{\textrm{eff}})+\textrm{vec}(\mathbf{U}) \\ = & P_{\textrm{Tx}}\left(\mathbf{W}_{1}^{T} \otimes \mathbf{f}^{T} \otimes \mathbf{C}^{H}\right) \left( \mathbf{A}(\bm{\varphi})^{*} \otimes \mathbf{A}(\bm{\theta}_{\textrm{BR}})^{*} \otimes \mathbf{A}(\bm{\phi}_{\textrm{RU}})  \right) \\  & \textrm{vec} \left(\textrm{diag}(\bm{\rho}_{\textrm{BR}}) \otimes \textrm{diag}(\bm{\rho}_{\textrm{RU}}) \right) + \mathbf{u} \in \mathbb{C}^{BM_{\textrm{U}} \times 1}.
\end{split}
\end{equation}
Here, $\mathbf{u}=\textrm{vec}(\mathbf{U})$ and $\mathbf{u} \sim \mathcal{CN}\left( \mathbf{0}_{BM_{\textrm{U}}},(\sigma^{2}P_{\textrm{Tx}}/D)\mathbf{I}_{BM_{\textrm{U}}} \right)$. Note that (\ref{3Dvec}) can be derived by Property 3 in Section~\ref{BTSec}.
Letting $\mathbf{z}_{\textrm{3D}}$ denotes $P_{\textrm{Tx}}\textrm{vec}(\mathbf{H}_{\textrm{eff}})$, $\bm{\Upsilon}$ can be rewritten as 
\begin{equation}
    \begin{split}
    \bm{\Upsilon}= \left(\mathbf{W}_{1}^{T} \otimes \mathbf{f}^{T} \otimes \mathbf{C}^{H}\right) \mathbf{z}_{\textrm{3D}} + \mathbf{u} \in \mathbf{C}^{BM_{\textrm{U}} \times 1}.  
    \end{split}
\end{equation}
To simplify the representation of following equations, we define $\mathbf{R}$ as $\mathbf{W}_{1}^{T} \otimes \mathbf{f}^{T} \otimes \mathbf{C}^{H}$ and $\mathcal{M}$ as $M_{\textrm{B}}M_{\textrm{R}}M_{\textrm{U}}$.

To define the atomic norm of $\mathbf{z}_{\textrm{3D}}$, the atomic set $\mathcal{A}_{\textrm{3D}}$ can be given by 
\begin{equation}
\begin{split}
    \mathcal{A}_{\textrm{3D}}=&\left\{ \mathbf{a}_{1}(\theta_{1})^{*} \otimes \mathbf{a}_{2}(\theta_{2})^{*} \otimes \mathbf{a}_{3}(\theta_{3}) \in \mathbb{C}^{\mathcal{M} \times 1}: \right. \\ & \quad \left. 0^{\circ} < \theta_{1},\theta_{2},\theta_{3} < 180^{\circ}  \right\},        
\end{split}
\end{equation}
where $\mathbf{a}_{1}(\theta_{1})^{*} \otimes \mathbf{a}_{2}(\theta_{2})^{*} \otimes \mathbf{a}_{3}(\theta_{3})$ is an atom for $\mathbf{z}_{\textrm{3D}}$, and $\mathbf{z}_{\textrm{3D}}$ can be represented as a linear combination of these atoms. 
$\mathcal{A}_{\textrm{3D}}$ is a set of 3D atoms defined in (\ref{NDatom}) and belongs to the SMV case.
The atomic norm for $\mathbf{z}_{\textrm{3D}}$, $\lVert \mathbf{z}_{\textrm{3D}} \rVert_{\mathcal{A}_{\textrm{3D}}}$ can be given by follows.
\begin{equation}
    \begin{split}   
    \lVert \mathbf{z}_{\textrm{3D}} \rVert_{\mathcal{A}_{\textrm{3D}}}=&\min_{\mathbf{T}_{\textrm{3D}},p} \;  \frac{1}{2 \mathcal{M}} \textrm{Tr}(\mathbf{T}_{\textrm{3D}})+ \frac{1}{2} p \\ &\; \textrm{s.t.}
    \begin{bmatrix}
    \mathbf{T}_{\textrm{3D}} & \mathbf{z}_{\textrm{3D}} \\
    \mathbf{z}_{\textrm{3D}}^{H} & p
    \end{bmatrix} \succeq 0.
    \end{split}
\end{equation}
Here, $\mathbf{T}_{\textrm{3D}}$ denotes 3D Toeplitz matrix and comprises of multiple 2D Toeplitz matrices. A structure of $\mathbf{T}_{\textrm{3D}}$ is given in~(\ref{3DToep}).
In~(\ref{3DToep}), $\mathbf{T}^{0}_{\textrm{2D}}$ denotes $M_{\textrm{B}}M_{\textrm{U}} \times M_{\textrm{B}}M_{\textrm{U}}$ 2-level Toeplitz matrix whose structure follows that of (\ref{2Dstructure}). 
However, the structure of $\mathbf{T}^{i}_{\textrm{2D}}$ for $i=1,\ldots,M_{\textrm{R}}$ is different from that of (\ref{2Dstructure}), where the structure of $\mathbf{T}^{i}_{\textrm{2D}}$ for $i=1,\ldots,M_{\textrm{R}}$ can be given by  
\begin{equation}\label{2Dstructure_nsym}
    \begin{split}
    \mathbf{T}^{i}_{\textrm{2D}}= 
    \begin{bmatrix}
    \mathbf{T}_{0} & \mathbf{T}_{-1} & \cdots & \mathbf{T}_{-M_{\textrm{B}}+1} & \mathbf{T}_{-M_{\textrm{B}}} \\
    \mathbf{T}_{1} & \mathbf{T}_{0} & \mathbf{T}_{-1} & \ddots & \mathbf{T}_{-M_{\textrm{B}}+1} \\
    \vdots & \ddots & \ddots & \ddots & \vdots \\
    \mathbf{T}_{M_{\textrm{B}}-1} & \ddots & \mathbf{T}_{1} & \mathbf{T}_{0} & \mathbf{T}_{-1} \\
    \mathbf{T}_{M_{\textrm{B}}} & \mathbf{T}_{M_{\textrm{B}}-1} & \cdots & \mathbf{T}_{1} & \mathbf{T}_{0}
    \end{bmatrix},\\
    \textrm{for}\; i=1,\ldots,M_{\textrm{R}}.
    \end{split}
\end{equation}
Note that Toeplitz matrices are not placed symmetrically as in (\ref{2Dstructure}).

\begin{figure*}[b]
    \begin{equation}\label{3DToep}
    \begin{aligned}
    \mathbf{T}_{\textrm{3D}}= 
    \begin{bmatrix}
    \mathbf{T}_{\textrm{2D}}^{0} & (\mathbf{T}_{\textrm{2D}}^{1})^{H} & \cdots & (\mathbf{T}_{\textrm{2D}}^{M_{\textrm{R}}-1})^{H} & (\mathbf{T}_{\textrm{2D}}^{M_{\textrm{R}}})^{H} \\
    \mathbf{T}_{\textrm{2D}}^{1} & \mathbf{T}_{\textrm{2D}}^{0} & (\mathbf{T}_{\textrm{2D}}^{1})^{H} & \ddots & (\mathbf{T}_{\textrm{2D}}^{M_{\textrm{R}}-1})^{H} \\
    \vdots & \ddots & \ddots & \ddots & \vdots \\
    \mathbf{T}_{\textrm{2D}}^{M_{\textrm{R}}-1} & \ddots & \mathbf{T}_{\textrm{2D}}^{1} & \mathbf{T}_{\textrm{2D}}^{0} & (\mathbf{T}_{\textrm{2D}}^{1})^{H} \\
    \mathbf{T}_{\textrm{2D}}^{M_{\textrm{R}}} & \mathbf{T}_{\textrm{2D}}^{M_{\textrm{R}}-1} & \cdots & \mathbf{T}_{\textrm{2D}}^{1} & \mathbf{T}_{\textrm{2D}}^{0}
    \end{bmatrix} \in \mathbb{C}^{\mathcal{M} \times \mathcal{M}}.
\end{aligned}
\end{equation}
\end{figure*}

\begin{table*}[t]
  \begin{center}
    \captionsetup{justification=centering}
    \caption{Summary of ANM-based channel estimation algorithms for RIS-aided MIMO systems}
    \label{tab:per}
    \begin{tabular}{|c||c|c|c|}
    \hline
     $\;$ & Non location-aware & \multicolumn{2}{c|}{Location-aware} \\ \cline{2-4}
     $\;$ & 1D ANM & 2D ANM & 3D ANM \\ \hline \hline
     Beam search at BS & Exhaustive search & Direct steering towards RIS & Direct steering towards RIS \\ \hline
     RIS beamwidth adaptation & O & X & O \\ \hline
     \makecell{Number of training symbols\\(Beam training overhead)} & $BP_{\textrm{B}}P_{\textrm{U}}$ & $M_{\textrm{R}}P_{\textrm{U}}$ & $BP_{\textrm{U}}$ \\ \hline
     Complexity & $\mathcal{O}\left((M_{\textrm{B}}M_{\textrm{U}}+M_{\textrm{R}})^{3.5}\right)$ & $\mathcal{O}\left((M_{\textrm{B}}M_{\textrm{U}}+M_{\textrm{R}})^{3.5}\right)$ & $\mathcal{O}\left((M_{\textrm{B}}M_{\textrm{R}}M_{\textrm{U}}+1)^{3.5}\right)$ \\ \hline
  \end{tabular}
  \end{center}
\end{table*}

An equation that estimates $\mathbf{z}_{\textrm{3D}}$ can be given by
\begin{equation}\label{opt3D}   
    \hat{\mathbf{z}}_{\textrm{3D}}=\argmin_{\bar{\mathbf{z}}_{\textrm{2D}}} \;  \tau_{\textrm{3D}} \lVert \bar{\mathbf{z}}_{\textrm{3D}} \rVert_{\mathcal{A}_{\textrm{3D}}} + \frac{1}{2} \lVert \bm{\Upsilon}-\mathbf{R}\bar{\mathbf{z}}_{\textrm{3D}} \rVert_{2}^{2},
\end{equation}
where $\hat{\mathbf{z}}_{\textrm{3D}}$ and $\bar{\mathbf{z}}_{\textrm{3D}}$ respectively denote the estimate of $\mathbf{z}_{\textrm{3D}}$ and the variable for estimation of $\mathbf{z}_{\textrm{3D}}$. 
For the SMV case, a regularization parameter $\tau_{\textrm{3D}}$ can be set as in~\cite{6560426}, considering the covariance of each element of $\mathbf{u}$.
\begin{equation}\label{tau3}   
    \tau_{\textrm{3D}}=\frac{\sigma\sqrt{P_{\textrm{Tx}}}}{\sqrt{D}}\Big(1+\frac{1}{\log \mathcal{M}} \Big) \sqrt{ \mathcal{M} \log \mathcal{M} + \mathcal{M} \log\left(4 \pi \log \mathcal{M} \right)}.
\end{equation}
Finally, (\ref{opt3D}) can be fully unfolded as
\begin{equation}\label{3Dunfold} 
    \begin{split}
    &\{\hat{\mathbf{T}}_{\textrm{3D}},\hat{p},\hat{\mathbf{z}}_{\textrm{3D}} \}=\\ & \argmin_{\mathbf{T}_{\textrm{3D}},p,\bar{\mathbf{z}}_{\textrm{2D}}} \;  \frac{\tau_{\textrm{3D}}}{2 \mathcal{M}} \textrm{Tr}(\mathbf{T}_{\textrm{3D}})+ \frac{\tau_{\textrm{3D}}}{2} p + \frac{1}{2} \lVert \bm{\Upsilon}-\mathbf{R}\bar{\mathbf{z}}_{\textrm{3D}} \rVert_{2}^{2} \\ &\; \textrm{s.t.}
    \begin{bmatrix}
    \mathbf{T}_{\textrm{3D}} & \bar{\mathbf{z}}_{\textrm{3D}} \\
    \bar{\mathbf{z}}_{\textrm{3D}}^{H} & p
    \end{bmatrix} \succeq 0. 
    \end{split}
\end{equation}
$\mathbf{H}_{\textrm{eff}}$ can be estimated by reshaping a vector $\hat{\mathbf{z}}_{\textrm{3D}}$ to $M_{\textrm{B}}M_{\textrm{U}} \times M_{R}$ matrix. 

A computational complexity of (\ref{3Dunfold}) can be derived by a theorem in~\cite{SDPcomp} after transcribing (\ref{3Dunfold}) into a general SDP form. 
If an influence of a stopping criteria for the optimization is not considered, $\mathcal{O}\left( (M_{\textrm{B}}M_{\textrm{R}}M_{\textrm{U}}+1)^{0.5} \right)$ iterations are required, where each iteration costs $\mathcal{O}\left( (M_{\textrm{B}}M_{\textrm{R}}M_{\textrm{U}}+1)^{3} \right)$.
Thus, the computational complexity of the location-aware channel estimation via 3D ANM is $\mathcal{O}\left( (M_{\textrm{B}}M_{\textrm{R}}M_{\textrm{U}}+1)^{3.5} \right)$.
Characteristics of three ANM-based channel estimation algorithms are summarized in Table~\ref{tab:per}.  

\section{Simulation Results and Discussions}\label{simulation}
\begin{figure}[t]
    \includegraphics[width=0.97\columnwidth]{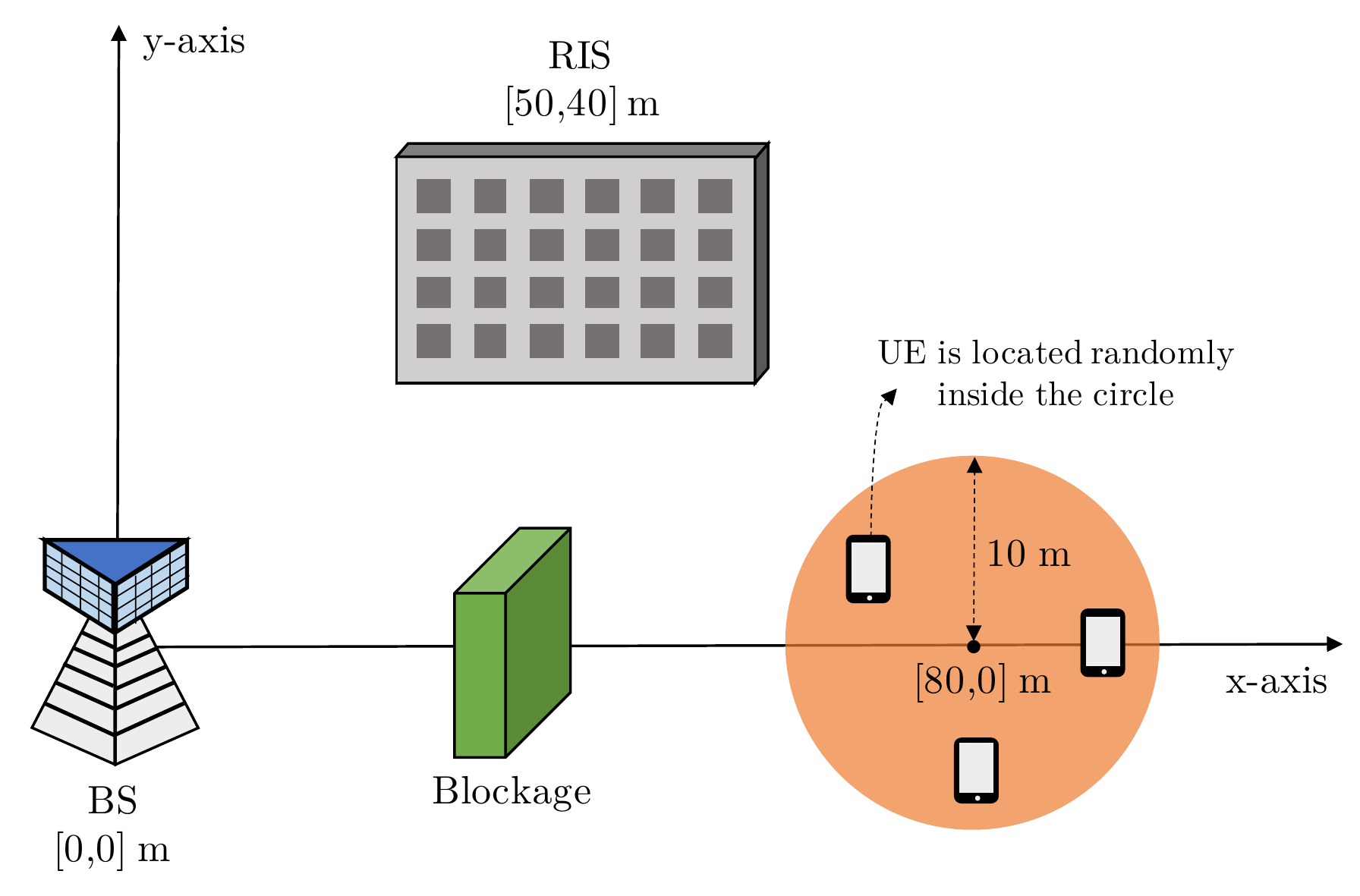}
    \caption{The location of the BS, the RIS, and the UE throughout simulation.}
    \label{SimulEnv}
\end{figure}

\subsection{Simulation Environments}
For the default simulation setting, $M_{\textrm{B}}$, $M_{\textrm{R}}$, and $M_{\textrm{U}}$ are respectively set to 4, 16, and 4, and $N_{\textrm{B}}$ and $N_{\textrm{U}}$ are respectively set to 2 and 2. 
The number of snapshots of pilot signal, $D$ is $100$. 
Precoding matrices and combining matrices are set as parts of scaled DFT matrices. Thus, $\mathbf{F}$ and $\mathbf{C}$ are respectively $\bm{\Psi}_{M_{\textrm{B}}}/\sqrt{M_{\textrm{B}}}$ and $\bm{\Psi}_{M_{\textrm{U}}}/\sqrt{M_{\textrm{U}}}$.
The transmission power is $30$ dBm so that $P_{\textrm{Tx}}=1000$. The noise power is $-100$ dBm so that $\sigma=10^{-5}$. 
For implementation of \cite{9354904}, a size of discretized grid is set to $360$ so that the angular domain is divided by $0.5^{\circ}$.

The location of the BS, the RIS, and the UE are set as in Fig.~\ref{SimulEnv}. The BS is located in $[0,0]$ m, and the RIS is located in $[50,40]$ m. If not noted, the UE is randomly located in the area of which the center is $[80,0]$ m and the radius is $10$ m. 
The BS, the RIS, and the UE are oriented in parallel.
Since there is only LoS path between the BS and the RIS, $\phi^{1}_{\textrm{BR}}$ and $\theta^{1}_{\textrm{BR}}$ are set as follows.
\begin{equation}
    \phi^{1}_{\textrm{BR}}=\theta^{1}_{\textrm{BR}}=\textrm{tan}^{-1}\left(\frac{40}{50} \right) = 38.66^{\circ}.
\end{equation}
The magnitude of $\alpha^{1}_{\textrm{BR}}$ is determined based on the distance between the BS and the RIS, the carrier frequency, and the path loss exponent. 
Here, the carrier frequency is $28$ GHz, and the path loss exponent is $2.1$. 
The phase of $\alpha^{1}_{\textrm{BR}}$ is randomly set between $0$ and $2 \pi$.

To model the RIS-to-UE channel, the 28 GHz statistical channel model in~\cite{7501500} is employed. The elevation of the RIS-to-UE AoD/AoA is not considered since the RIS and the UE are using ULAs. Also, we assume that the time cluster comprises of one signal path.
$L_{\textrm{RU}}$ is randomly set between $1$ and $6$, and the LoS path between the RIS and the UE exists if the distance between the RIS and the UE is less than $60$ m.
The channel gains and the RIS-to-UE AoDs/AoAs are determined randomly based on~\cite{7501500}. 
However, if the LoS path exists between the RIS and the UE, $\phi^{1}_{\textrm{RU}}$ and $\theta^{1}_{\textrm{RU}}$ are set as
\begin{equation}
    \phi^{1}_{\textrm{RU}}=\theta^{1}_{\textrm{RU}}=\textrm{tan}^{-1}\left(\frac{40-\textrm{y}_{\textrm{UE}}}{\textrm{x}_{\textrm{UE}}-50} \right),
\end{equation}
where $\textrm{x}_{\textrm{UE}}$ and $\textrm{y}_{\textrm{UE}}$ respectively denote the x-coordinate and the y-coordinate of the UE.

The SNR is defined as
\begin{equation}\label{SNR}
    \textrm{SNR}=10 \log_{10} \frac{P_{\textrm{Tx}} \left| \left( \sum_{l=1}^{L_{\textrm{RU}}} \alpha_{\textrm{RU}}^{l} \right) \left( \sum_{l=1}^{L_{\textrm{BR}}} \alpha_{\textrm{BR}}^{l} \right) \right|^{2}}{\sigma^{2}}  (\textrm{dB}).
\end{equation}
The SNR defined in~(\ref{SNR}) is a ratio of signal power and noise power when $M_{\textrm{B}}=M_{\textrm{R}}=M_{\textrm{U}}=1$. 
Note that the SNR defined in~(\ref{SNR}) does not depend on precoding matrix, combining matrix, and RIS control matrix. 
The normalized mean square error (NMSE) is defined as
\begin{equation}
    \textrm{NMSE}=\frac{1}{Q}\sum_{q=1}^{Q} \frac{\lVert \hat{\mathbf{H}}^{q}_{\textrm{eff}} - \mathbf{H}^{q}_{\textrm{eff}} \rVert^{2}_{\textrm{F}}}{\lVert \mathbf{H}^{q}_{\textrm{eff}} \rVert^{2}_{\textrm{F}}},        
\end{equation}
where $Q$ is the number of Monte Carlo trials for NMSE calculation and is set to 300.
$\hat{\mathbf{H}}^{q}_{\textrm{eff}}$ and $\mathbf{H}^{q}_{\textrm{eff}}$ respectively denote the estimated effective cascaded channel and the actual effective cascaded channel on the $q$-th Monte Carlo trial.
For computation, Intel CPU i5-7500 (3.40 GHz) and 16 GB RAM are used, and CVX~\cite{cvx} is used to solve (\ref{final2}), (\ref{2Dunfold}), (\ref{3Dunfold}), and the algorithm in~\cite{he2021channel}.

\begin{figure*}[t]
    \begin{center}
    \captionsetup[subfigure]{justification=centering}
    \begin{subfigure}[t]{0.65\columnwidth}
        \includegraphics[width=\columnwidth,center]{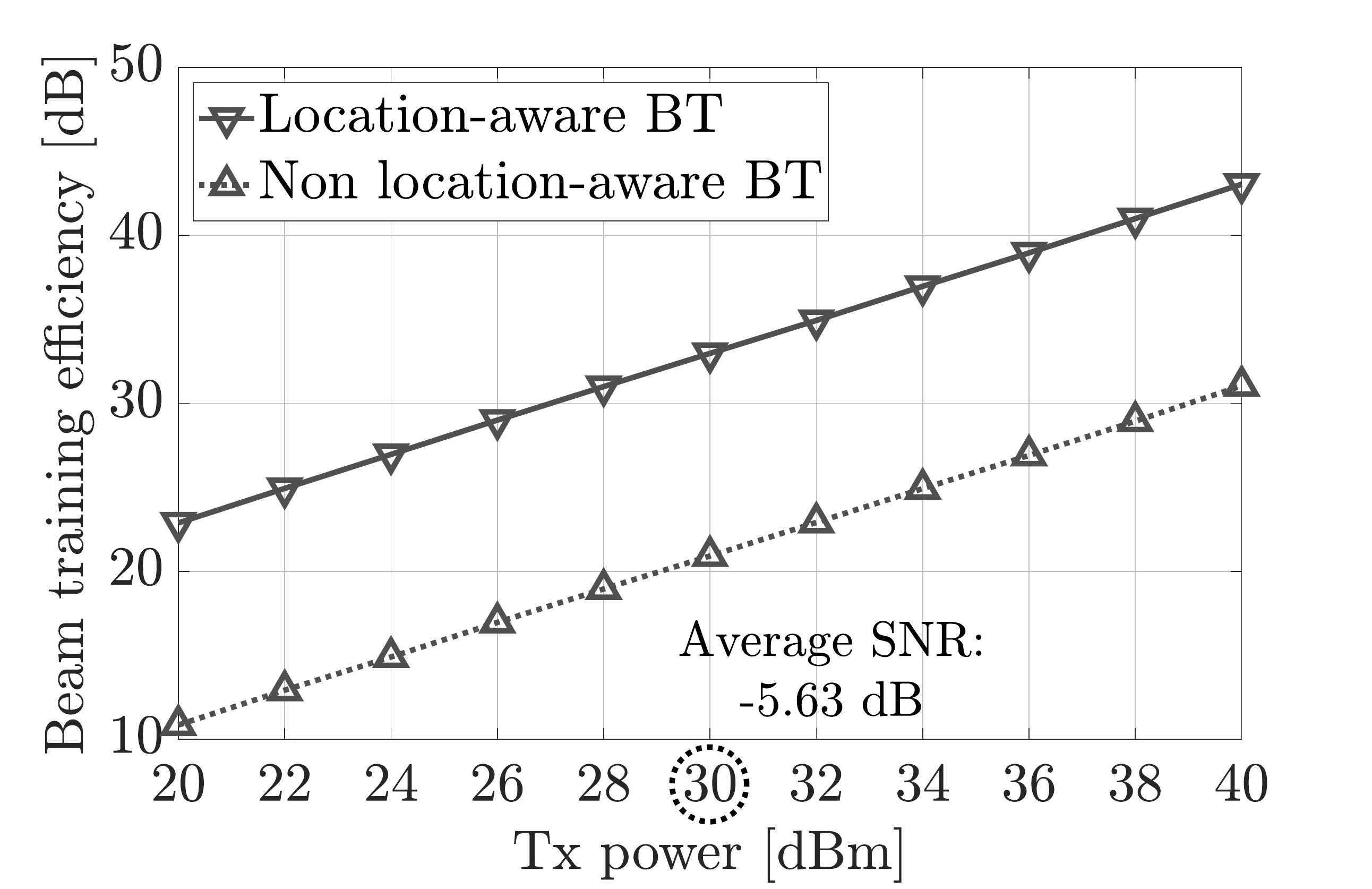}
        \caption{Beam training efficiency versus Tx power}\label{1st}
    \end{subfigure}
    \begin{subfigure}[t]{0.65\columnwidth}
        \includegraphics[width=\columnwidth,center]{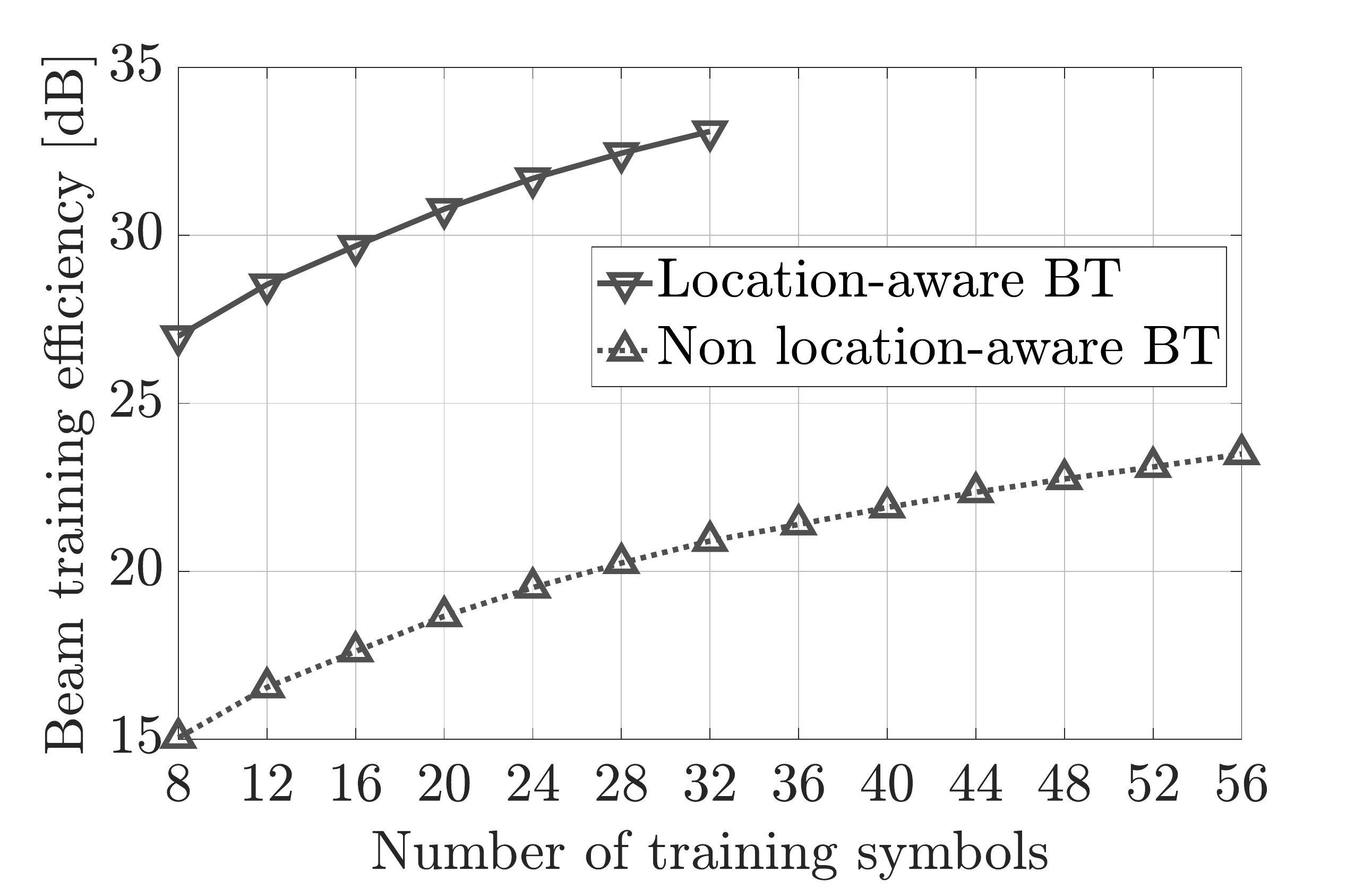}
        \caption{Beam training efficiency versus number of training symbols}\label{2nd}
    \end{subfigure}
    \begin{subfigure}[t]{0.65\columnwidth}
        \includegraphics[width=\columnwidth,center]{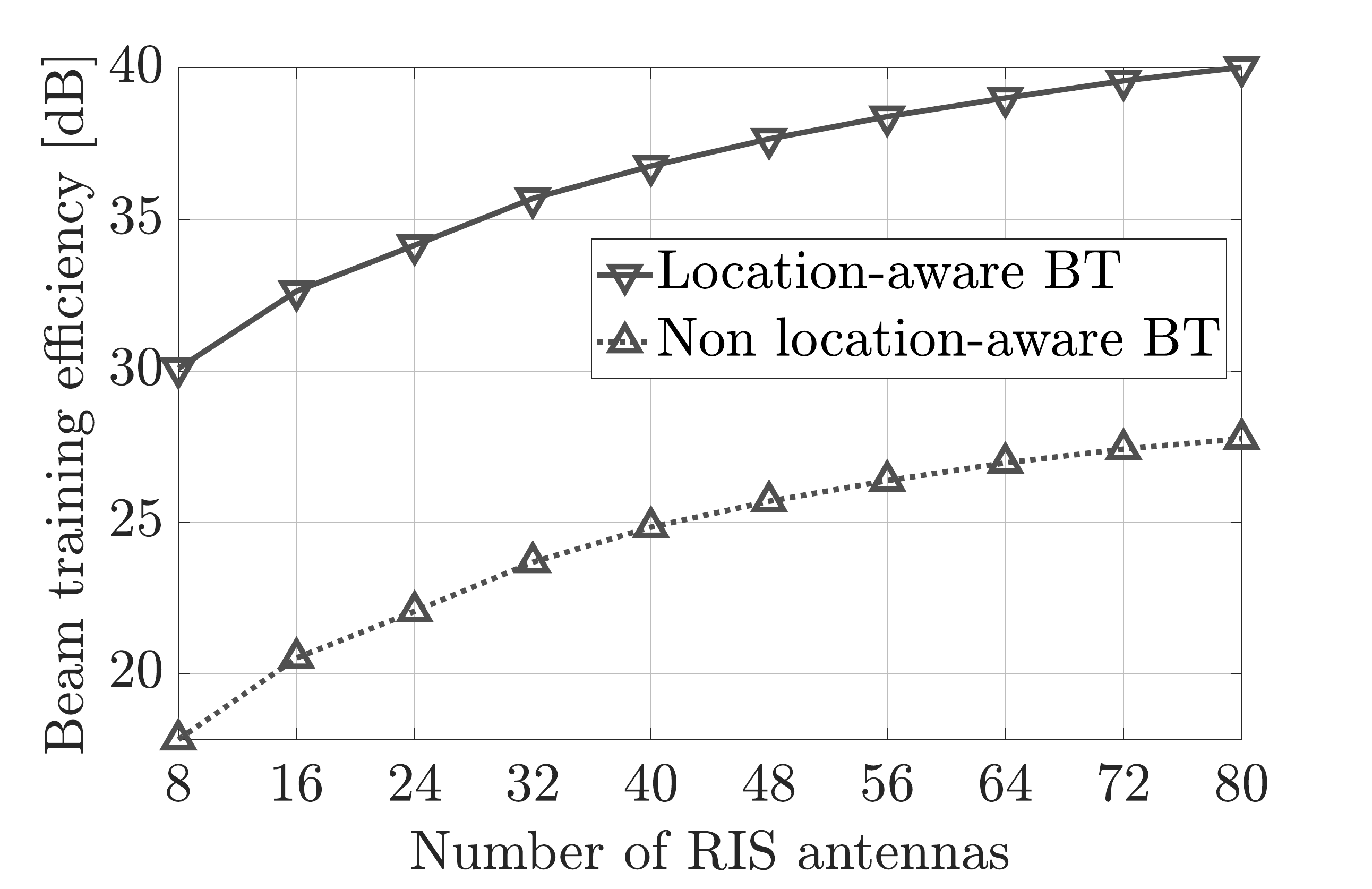}
        \caption{Beam training efficiency versus number of RIS antennas}\label{3rd}
    \end{subfigure}
    \caption{Beam training efficiency according to the use of location information. In (a), the number of training symbols is set to $32$ for both beam training methods. In (c), the number of training symbols is set to twice the number of the RIS antennas.}\label{Loc} 
    \end{center}
\end{figure*}

\subsection{Benefit and Limitation of Location-aware Beam Training}
In this subsection, the benefit and the limitation of using location are presented by analyzing the beam training efficiency defined in (\ref{eSNR_NLA}) and (\ref{eSNR_LA}). 
Three subfigures in Fig.~\ref{Loc} respectively present the beam training efficiency versus the transmission power, the number of training symbols, and the number of RIS antennas.
The number of training symbols is set to $32$ for both beam training methods in Fig.~\ref{1st}.
In Fig.~\ref{2nd}, the maximum number of training symbols for the location-aware beam training is $32$, and the number of training symbols is set to twice the number of the RIS antennas in Fig.~\ref{3rd}.
In Fig.~\ref{Loc}, the beam training efficiency increases by $12$ dB when using location. Main reasons for the significant improvement can be summarized as follows:
\begin{itemize}
    \item Since the BS steers the beam towards the RIS when using location, the power of the pilot signal received during the beam training increases. Also, the BS focuses the transmission power to the single beam.
    \item If the number of training symbols is identical for both beam training methods, the RIS can use sharper reflect beams when performing the location-aware beam training. 
\end{itemize}
According to the results in Fig.~\ref{Loc}, the location-aware channel estimation is expected to be superior to the non location-aware channel estimation when there is no location error.

\begin{figure}[t]
    \captionsetup[subfigure]{justification=centering}
        \includegraphics[width=1\columnwidth]{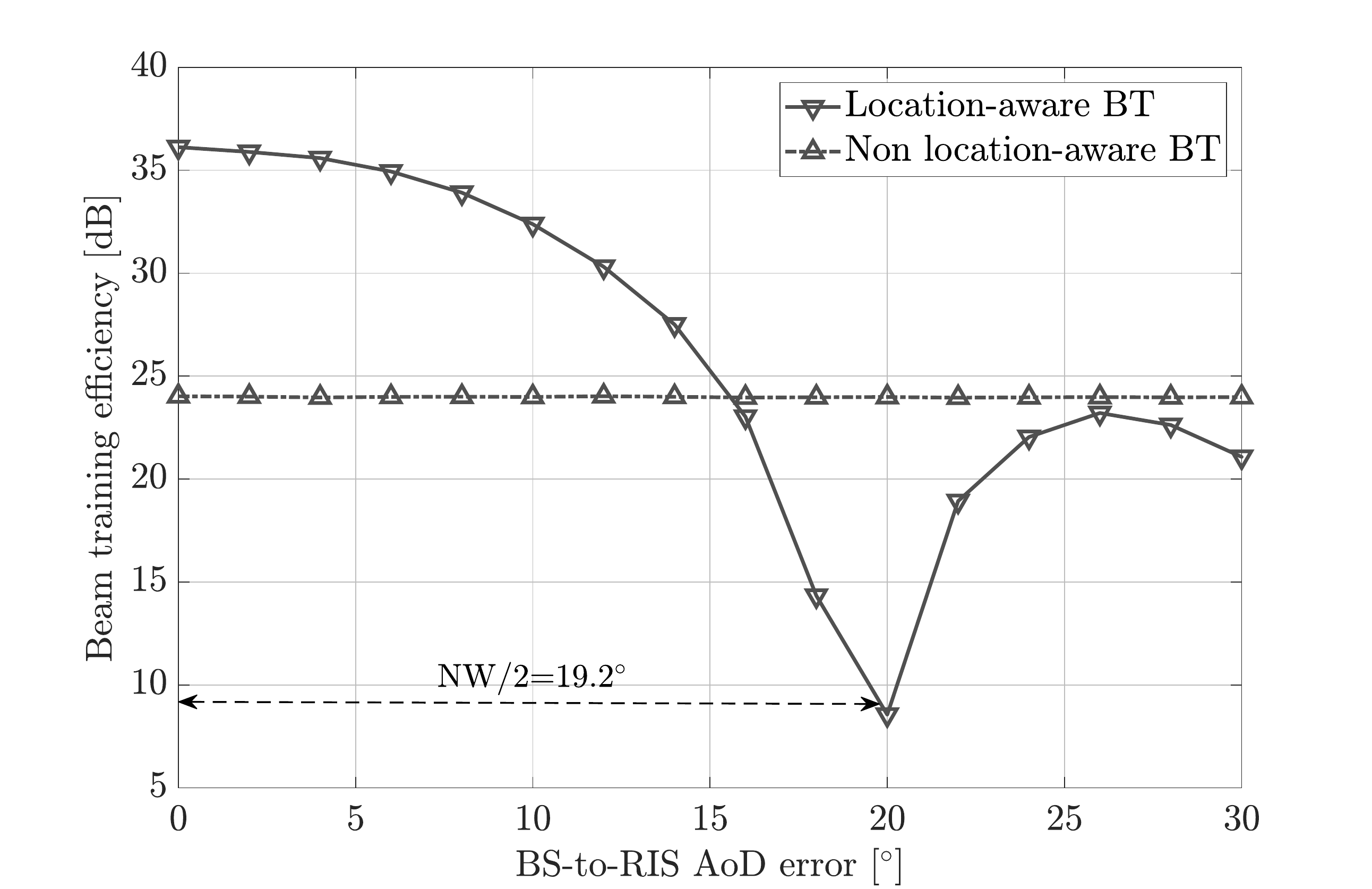}
        \caption{Beam training efficiency according to the BS-to-RIS AoD error. $M_{\textrm{B}}=8$, and the number of training symbols is set to $32$ for both beam training methods.}
    \label{EffLoc}
\end{figure}

Although the location of the BS and the RIS are expected to be accurate when the BS and the RIS are immobile, there are scenarios that the location can be erroneous.
For example, the location of the RIS can be inaccurate when using mobile RIS, which has been studied in seminal works~\cite{9356531,AER3}.
In this scenario, the BS-to-RIS AoD does not match with the beam steering direction of the BS when performing location-aware beam training.
Fig.~\ref{EffLoc} shows the beam training efficiency according to the BS-to-AoD error, which is induced by the location error of the RIS. In this simulation, the BS-to-RIS distance and the RIS-to-AoD distance are respectively fixed to $\sqrt{4100}$ m and $50$ m. The number of training symbols is set to $32$ for both beam training methods, and $M_{\textrm{B}}=8$. 
In Fig.~\ref{EffLoc}, the beam training efficiency of the location-aware beam training becomes minimum when the BS-to-RIS AoD error reaches the half of the null-to-null beamwidth, where the half of the null-to-null beamwidth is the distance from the steering direction to the first null~\cite{ARRAY}.
On the other hand, the beam training efficiency of the non location-aware beam training is not affected by the BS-to-RIS AoD error.

\subsection{Analysis on Channel Estimation Performance}
In this subsection, we analyze the NMSE of the effective cascaded channel with respect to various factors that can affect the performance-transmission power, number of training symbols, number of RIS antennas, RIS-to-UE distance, and BS-to-RIS AoD error.
The former studies on the channel estimation for RIS-aided MIMO systems, \cite{9354904} and \cite{he2021channel}, are benchmarks for performance evaluation.
Note that \cite{9354904} and \cite{he2021channel} also employ the RIS beamwidth adaptation. 
The algorithm in \cite{9103231} is excluded from the comparison since the computation time becomes excessively high when reforming the algorithm to work under RIS-aided MIMO systems.

\begin{figure}[t]
    \captionsetup[subfigure]{justification=centering}
        \includegraphics[width=1\columnwidth]{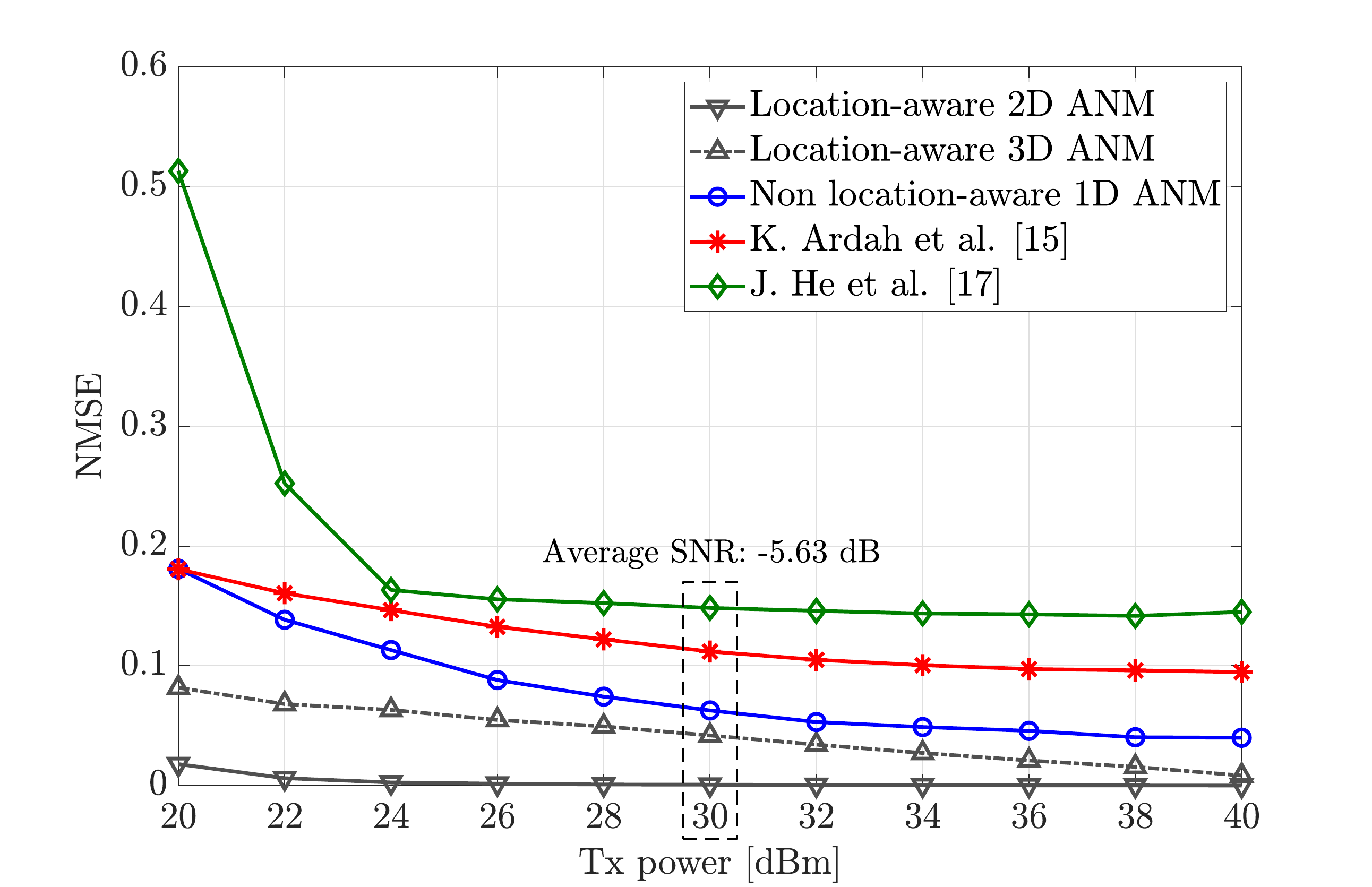}
        \caption{NMSE versus transmission power. The average SNR is $-5.63$ dB when the transmission power is $30$ dBm.}
    \label{ResultTx}
\end{figure}

Fig.~\ref{ResultTx} shows the NMSE according to the transmission power. The average SNR is $-5.63$ dB when the transmission power is $30$ dBm, and the average SNR is proportional to the transmission power.
In this simulation, the number of training symbols is set to $32$ except for the location-aware channel estimation via 3D ANM, where $16$ training symbols are used for the location-aware channel estimation via 3D ANM. 
This is because there is no need to employ the 3D ANM-based algorithm since the 2D ANM-based algorithm is available if using $32$ training symbols.
In Fig.~\ref{ResultTx}, the NMSE of the location-aware channel estimation via 2D ANM and 3D ANM are lower than those of the non location-aware channel estimation algorithms.
The NMSE of the 3D ANM-based algorithm is higher than that of the 2D ANM since the 3D ANM-based algorithm uses only $16$ training symbols. 
Still, the 3D ANM-based algorithm shows superior estimation accuracy over other non location-aware algorithms even when the beam training overhead is further reduced. 

\begin{figure}[!t]
    \includegraphics[width=1\columnwidth]{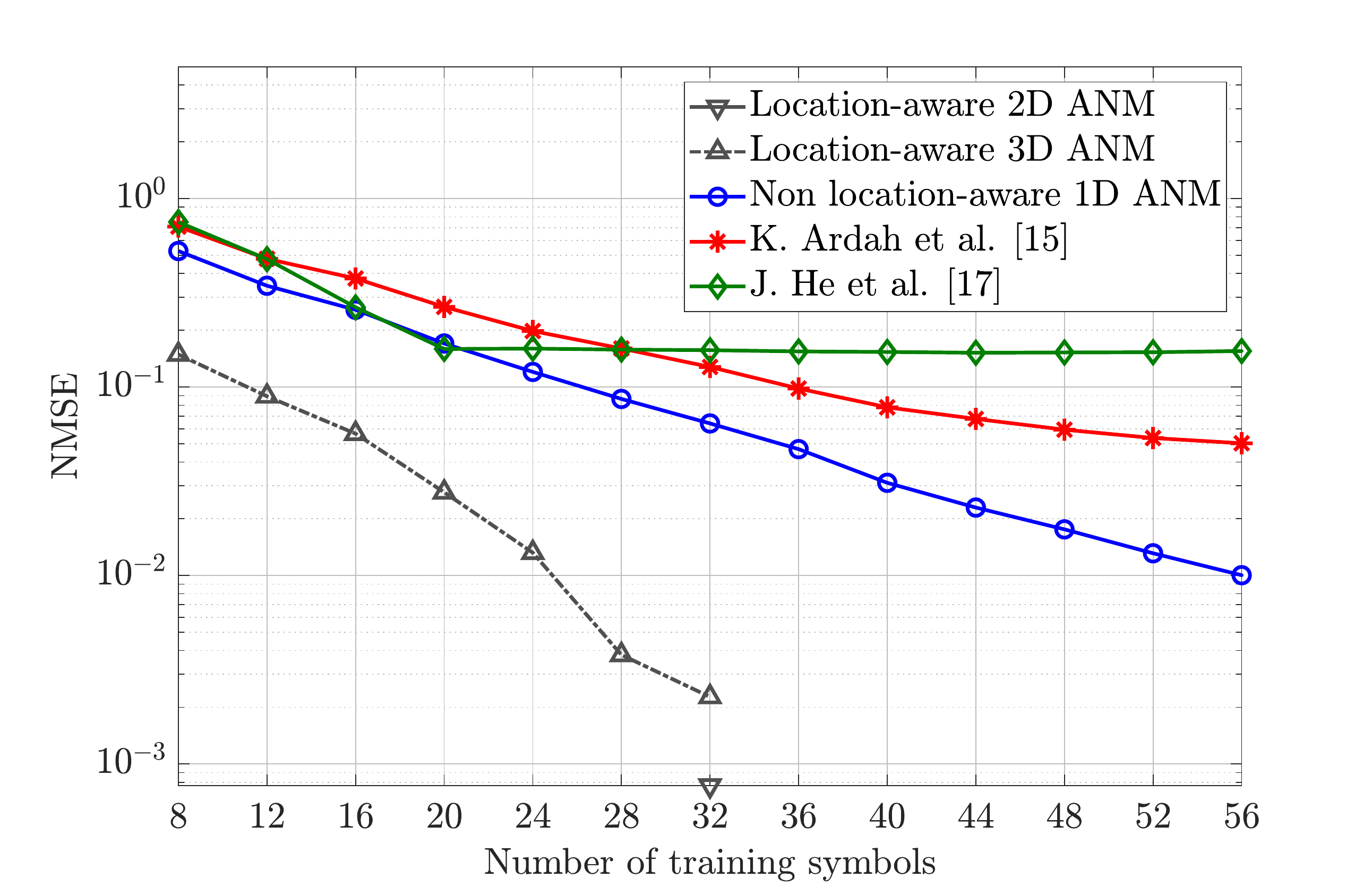}
    \caption{NMSE versus number of training symbols.}
    \label{ResultBT}
\end{figure}

Fig.~\ref{ResultBT} shows the NMSE according to the number of training symbols.  
The number of active RIS antennas is adjusted based on the number of training symbols. 
The only available number of training symbols for the location-aware channel estimation via 2D ANM is $32$, and the maximum number of training symbols for the location-aware channel estimation via 3D ANM is $32$.
Fig.~\ref{ResultBT} presents that the NMSE of all algorithms decrease as the number of training symbols increases, and the location-aware channel estimation is superior to the non location-aware channel estimation.
As shown in Fig.~\ref{2nd}, the beam training efficiency increases when using location if the number of training symbols is identical.
For this reason, the location-aware channel estimation algorithms achieve lower NMSE than others when consuming the same beam training overhead.
Among the non location-aware channel estimation algorithms, the 1D ANM-based channel estimation has the lowest NMSE.

\begin{figure}[t]
    \captionsetup[subfigure]{justification=centering}
    \begin{subfigure}[t]{1\columnwidth}
        \includegraphics[width=1\columnwidth]{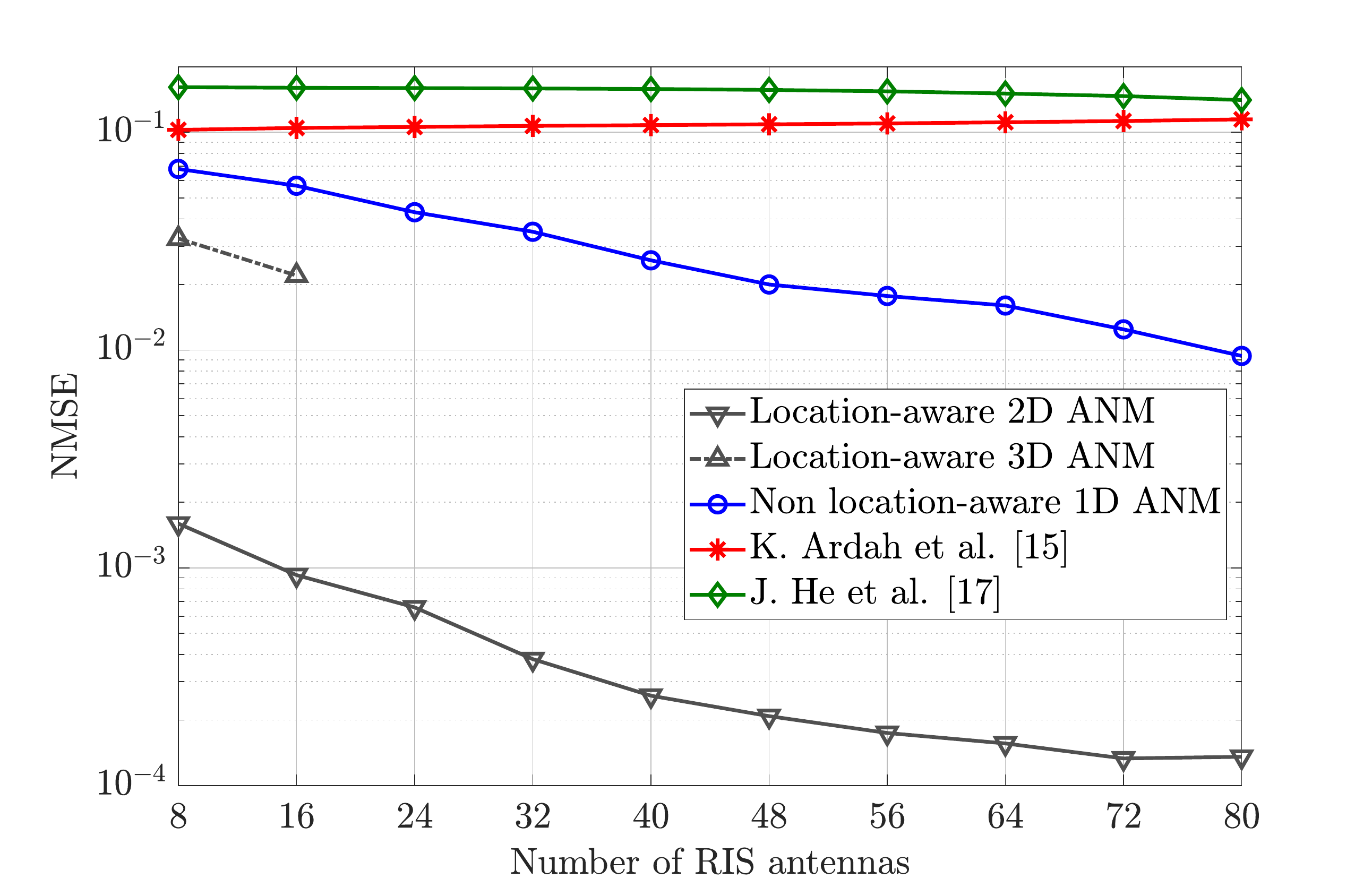}
        \caption{NMSE versus number of RIS antennas}\label{NMSERIS}
    \end{subfigure}
    \begin{subfigure}[t]{1\columnwidth}
        \includegraphics[width=1\columnwidth]{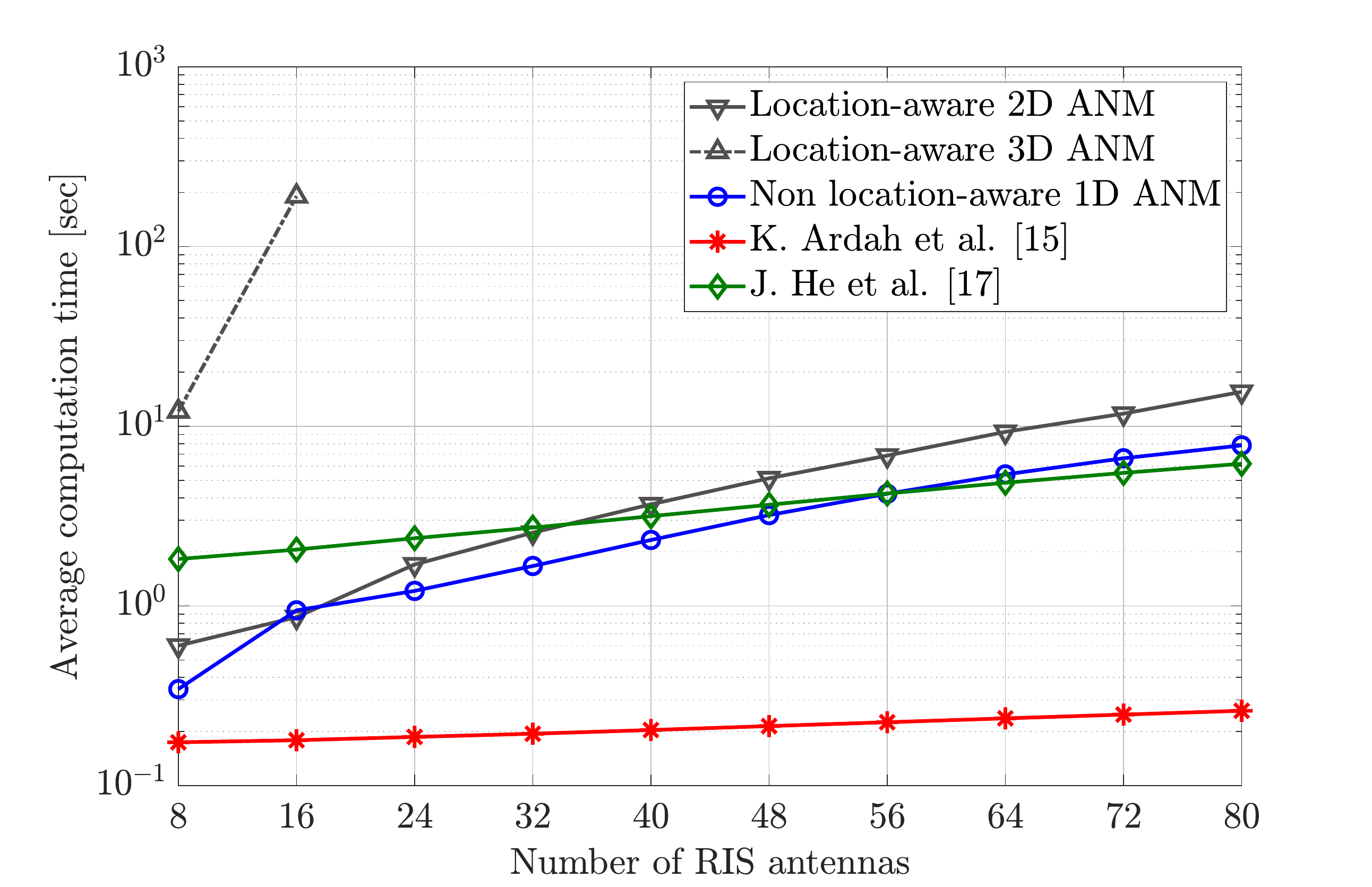}
        \caption{Average computation time versus number of RIS antennas}\label{TimeRIS}
    \end{subfigure}
    \caption{NMSE and average computation time according to the number of RIS antennas. Since the computing system cannot afford the 3D ANM-based algorithm if $M_{\textrm{R}}>16$, the analysis on the 3D ANM-based algorithm is omitted when $M_{\textrm{R}}>16$.}
    \label{ResultRIS}
\end{figure}

Fig.~\ref{ResultRIS} shows the NMSE and the average computation time according to the number of the RIS antennas.
Since the computing system cannot afford the 3D ANM-based algorithm if $M_{\textrm{R}}>16$, the analysis of the 3D ANM-based algorithm is omitted when $M_{\textrm{R}}>16$.
For all algorithms except the 3D ANM-based algorithm, the number of the training symbols is set to twice the number of the RIS antennas.
For the 3D ANM-based algorithm, the number of the training symbols equals to the number of the RIS antennas.
In Fig.~\ref{ResultRIS}, the NMSE of all algorithms are inversely proportional to the number of the RIS antennas except~\cite{9354904}.
In case of~\cite{9354904}, the AoD/AoA estimation accuracy does not improve with the number of the RIS antennas due to the grid-mismatch. 
If the number of the RIS antennas increases, the channel estimation accuracy becomes more dependent on the AoD/AoA estimation accuracy since even the small angle variation significantly affects the last elements of the steering vector.
In Fig.~\ref{TimeRIS}, the computation time of the 3D-ANM based algorithm is exceptionally high, where this result can be inferred by the complexity analysis in Table~\ref{tab:per}.
Other algorithms consume a similar level of the average computation time except for \cite{9354904}, which takes about $0.2$ seconds. 

\begin{figure}[t]
        \includegraphics[width=1\columnwidth]{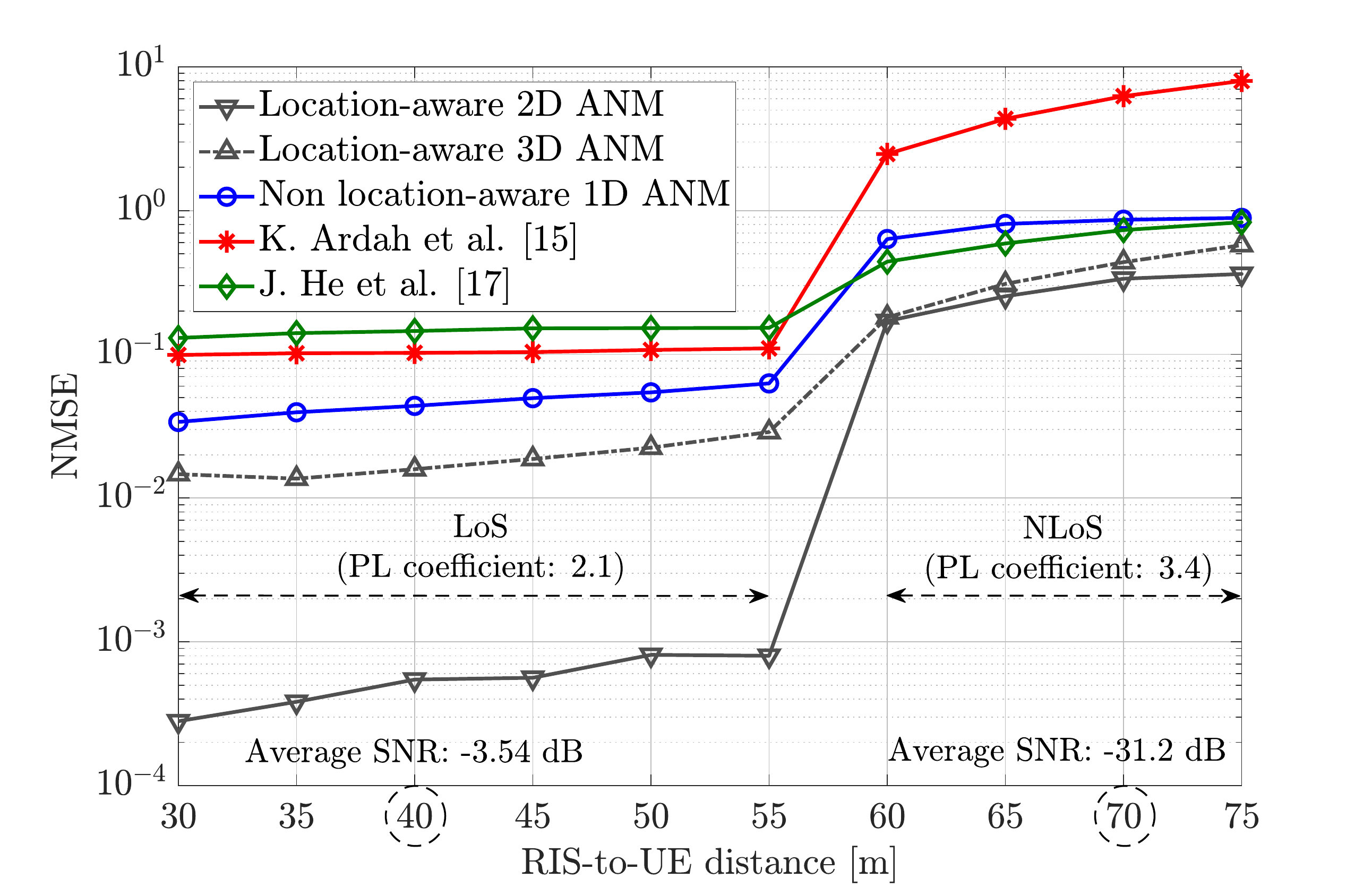}
        \caption{NMSE versus RIS-to-UE distance. The average SNR are $-3.54$ dB and $-31.2$ dB when the RIS-to-UE distance are $40$ m and $70$ m.}
    \label{ResultDis}
\end{figure}

Fig.~\ref{ResultDis} shows the NMSE according to the RIS-to-UE distance. 
In this simulation, the UE is located so that it is separated by a specified distance from the RIS. A number of training symbols is set to $32$ except for the location-aware channel estimation via 3D ANM, where the location-aware channel estimation via 3D ANM uses $16$ training symbols. 
When using the statistical mmWave channel model in~\cite{7501500}, the dominant LoS path between RIS and UE disappears if the RIS-to-UE distance goes beyond $60$ m, and the path loss coefficient for LoS environment and NLoS environment are respectively $2.1$ and $3.4$.
For this reason, the SNR changes abruptly around $60$ m.
In Fig.~\ref{ResultDis}, the non location-aware channel estimation algorithms fail under NLoS environment since the NMSE are around or beyond $1$. 
When the NMSE is around $1$, it means that elements of the estimated effective cascaded channel are near $0$. If the NMSE is beyond $1$, it means that the estimation is completely wrong.
On the other hand, the location-aware channel estimation algorithms frequently succeed in estimating the effective cascaded channel even under the NLoS environment, and this results in lower NMSE.  

\begin{figure}[!t]
    \includegraphics[width=1\columnwidth]{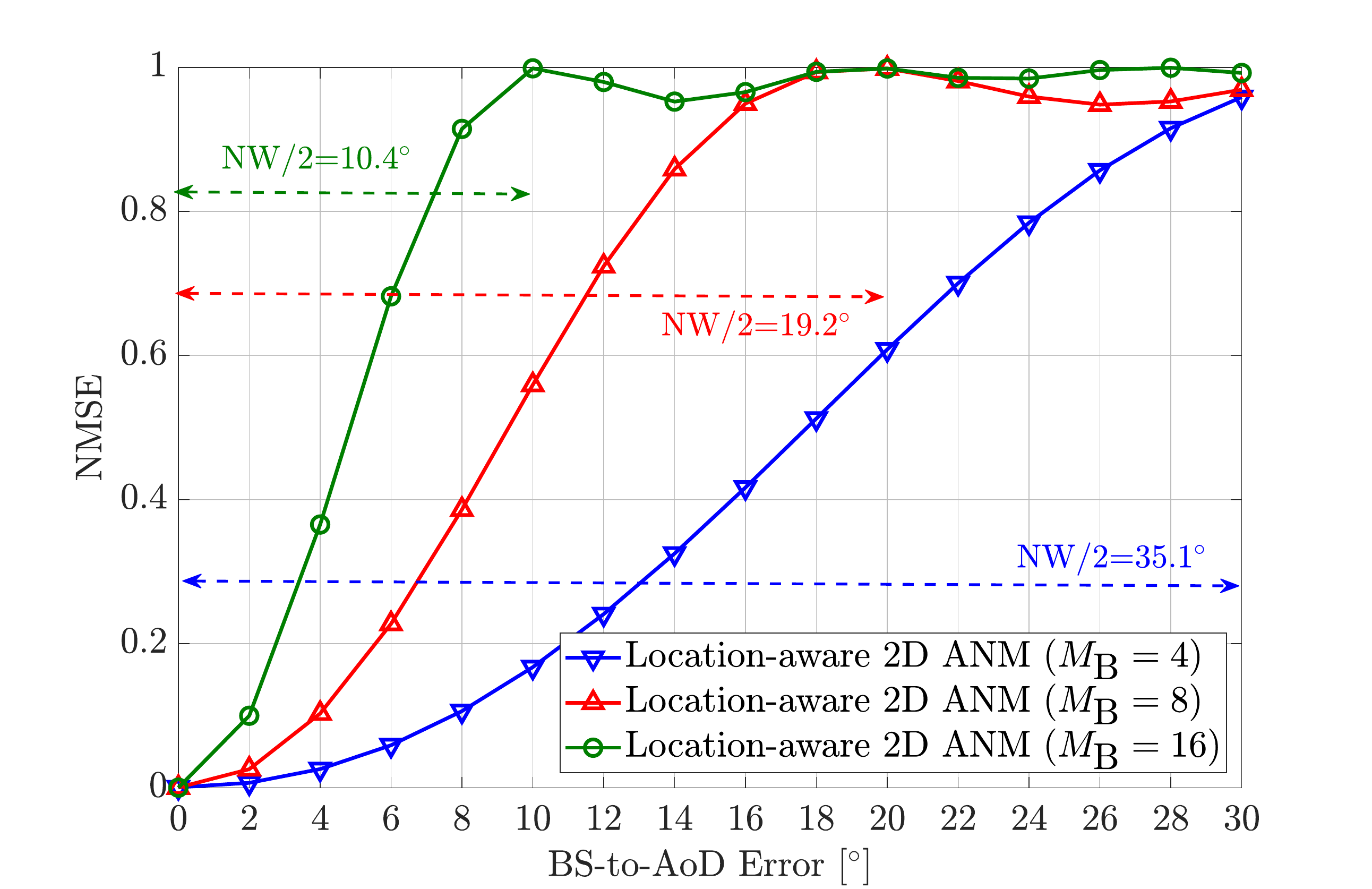}
    \caption{NMSE versus BS-to-RIS AoD error. Colored dotted lines denote the half of the null-to-null beamwidth according to the number of antennas on the BS.}
    \label{ResultLoc}
\end{figure}

Fig.~\ref{ResultLoc} shows the NMSE according to the BS-to-RIS AoD error, which is induced by the location error of the RIS. The BS-to-RIS distance and the RIS-to-UE distance are respectively fixed to $\sqrt{4100}$ m and $50$ m.
Colored dotted lines denote the half of the null-to-null beamwidth according to the number of antennas on the BS.  
Analysis of the location-aware channel estimation via 3D ANM is omitted since the computation time becomes excessive when $M_{\textrm{B}}>4$.
Fig.~\ref{ResultLoc} shows that the performance of the location-aware channel estimation degrades as the location error increases. 
To be more specific, the NMSE of the location-aware channel estimation becomes $1$ when the BS-to-RIS AoD reaches the null. 
This is because the strength of the pilot signal decreases when the BS-to-RIS AoD reaches closer to the null. 
Thus, the location-aware algorithms may not perform well when the location of the RIS is erroneous and the BS forms highly directional beam.    

\section{Conclusions}\label{conclusion}
In this paper, we propose a location-aware channel estimation based on the ANM for the RIS-aided MIMO systems.
With the location of the BS and the RIS, the beam training overhead at the BS is reduced by the beam steering towards the RIS.
Also, the RIS beamwidth adaptation reduces the beam training overhead at the RIS while ensuring the UE receives all the multipath components from the RIS.
When the BS steers the beam towards the RIS during beam training, the effective cascaded channel is estimated via either 2D ANM or 3D ANM.
Here, if the beam training overhead is further reduced by RIS beamwidth adaptation, the effective cascaded channel is represented as a linear combination of 3D atoms, hence is estimated via 3D ANM. 
The simulation results present that the location-aware channel estimation via 2D ANM and 3D ANM achieves superior estimation accuracy to benchmarks. Still, the location-aware channel estimation via 3D ANM is expected to be more useful if its computational complexity is reduced. Also, the location-aware channel estimation that are robust against the location error remains for further study.

\ifCLASSOPTIONcaptionsoff
  \newpage
\fi

\bibliographystyle{ieeetr}
\bibliography{reference}


\end{document}